\documentclass[12pt]{article}
\usepackage{amsmath,amsfonts,amssymb,graphicx,geometry,setspace,hyperref,natbib, multicol, booktabs, latexsym, bm, dashrule}
\usepackage{color, subcaption, multirow}
\title{Bayesian Scalar-on-Tensor Quantile Regression for Longitudinal Data on Alzheimer's Disease}
\author{Rongke Lyu\footnote{Department of Statistics, Rice University, Houston, TX, USA}, Marina Vannucci$^*$ and Suprateek Kundu\footnote{Department of Biostatistics,
MD Anderson Cancer Center, Houston, TX, USA}}
\date{}

\begin{document}

\maketitle

\begin{abstract}
As a general and robust alternative to traditional mean regression models, quantile regression avoids the assumption of normally distributed errors, making it a versatile choice when modeling outcomes such as cognitive scores that typically have skewed distributions. Motivated by an application to Alzheimer's disease data where the aim is to explore how brain-behavior associations change over time, we propose a novel Bayesian tensor quantile regression for high-dimensional longitudinal imaging data. The proposed approach distinguishes between effects that are consistent across visits and patterns unique to each visit, contributing to the overall longitudinal trajectory. A low-rank decomposition is employed on the tensor coefficients which reduces dimensionality and preserves spatial configurations of the imaging voxels. We incorporate multiway shrinkage priors to model the visit-invariant tensor coefficients and variable selection priors on the tensor margins of the visit-specific effects. For posterior inference,  we develop a computationally efficient Markov chain Monte Carlo sampling algorithm. Simulation studies reveal significant improvements in parameter estimation, feature selection, and prediction performance when compared with existing approaches. In the analysis of the Alzheimer's disease data, the flexibility of our modeling approach brings new insights as it provides a fuller picture of the relationship between the imaging voxels and the quantile distributions of the cognitive scores.
\end{abstract}


\section{Introduction}

Alzheimer's disease (AD) is a neurodegenerative disease of the brain that affects an estimated 7 million Americans aged 65 or older, with an estimated public health cost of around $360$ billion. This number could grow to 13.8 million by 2060 barring the development of medical breakthroughs \citep{better2023alzheimer}. Neuroimaging studies have emerged as an invaluable tool for identifying individuals susceptible to neurological and clinical conditions. By discerning subtle abnormalities in brain structure and function, researchers can now predict the risk of mental disorders (including AD and dementia) with greater precision. In particular, structural neuroimaging studies involving magnetic resonance imaging (MRI), have revealed the intricate patterns of atrophy in spatially distributed brain regions involved in memory and cognition, which is a hallmark of AD \citep{frenzel2020biomarker}. Neuroimaging features such as brain volume or cortical thickness (CT) derived from MRI scans can be used to provide quantitative assessments of structural atrophy and monitor disease progression over time, with CT emerging as a highly sensitive neuroimaging biomarker for AD  \citep{du2007different,weston2016presymptomatic}.
Progression to dementia due to AD involves multiple pathways of disease pathophysiology that impact cognition \citep{jack2018nia, jack2013tracking, jagust2018imaging}. Individuals who develop dementia follow a trajectory from a stage of normal cognition to Mild Cognitive Impairment (MCI) and subsequent dementia \citep{mckhann2011diagnosis, petersen2001current, sperling2011toward}. There has been very limited work on attempting to delineate gradual changes due to disease progression over visits, which is particularly critical given the current 2018 NIA-AA research framework that has transitioned to defining
AD as a continuum \citep{aisen2017path}. For instance, there is a lack of principled statistical approaches that can map spatially distributed longitudinal CT changes in the brain surface and borrow information across longitudinal brain images to model cognitive impairment. 

Standard neuroimaging approaches for risk prediction typically use informative summary measures or voxel-wise analysis \citep{rathore2017review}. These approaches are useful in terms of reducing the dimensions from high-dimensional images containing tens of thousands of voxels to a smaller subset of informative global or region-specific features. However, they result in potential information loss and do not respect the spatial information in the images. Scalar-on-image regression methods have been proposed that carefully account for the spatial configuration of hundreds of thousands of voxels, via basis representations of the images. These approaches may be considered as an extension of scalar-on-function models to functional imaging data, and involve methods such as functional principal components \citep{zipunnikov2011functional,feng2020bayesian}, as well as wavelet-based methods \citep{wang2014regularized,reiss2015wavelet,ma2024multi}. Separately, Markov random field (MRF) methods have been proposed \citep{lee2014spatial,smith2007spatial}. While they are based on solid theoretical foundations, MRF approaches and basis expansion methods typically involve a large number of parameters to model high-dimensional images that may induce the curse of dimensionality. To bypass these issues, scalar-on-tensor models have seen an increasing popularity for modeling image data \citep{guhaniyogi2017bayesian,zhou2013tensor,liu2023integrative,lyu2024bayesian}. Tensor-based approaches are attractive since they result in massive dimension reduction and they simultaneously preserve the spatial information in the images \citep{kundu2023bayesian}. 

The bulk of tensor modeling literature, including the above approaches, focuses on mean regression models involving normally distributed errors. However, the Gaussianity assumptions may not hold when modeling complex outcomes such as cognitive scores that typically have skewed distributions, see Figure \ref{adas-distribution} for an illustration of the outcome data used in our application. Additionally, when modeling cognitive outcomes as in our motivating application, it may be more meaningful to investigate the effect of imaging covariates on the outcome distribution, rather than solely focusing on the mean. One can achieve this type of analysis via quantile regression \citep{koenker1978regression,koenker2001quantile}, which offers a more robust approach than mean regression by handling outliers and accommodating deviations from normality. 
Bayesian quantile regression approaches, in particular, have seen a rich development over the past couple of decades. 
\cite{yu2001bayesian} proposed a Bayesian approach to quantile regression that uses asymmetric Laplace Distribution (ALD) priors. \cite{dunson2005approximate} proposed a substitution likelihood method to approximate Bayesian inferences on differences in quantiles and \cite{feng2015bayesian} proposed a Bayesian method that uses linear interpolation of the quantiles to approximate the likelihood. Among nonparametric Bayesian approaches, \cite{kottas2009bayesian} proposed a semiparametric approach based on Dirichlet process mixture models. \cite{kottas2007model} proposed an additive quantile regression framework with Gaussian process priors, that was later extended to allow for partially observed responses in a Tobit quantile regression framework by \cite{taddy2010bayesian}. \cite{reich2013bayesian} proposed a Bayesian quantile regression model for censored data. Additionally, \cite{tokdar2012simultaneous} proposed a semi-parametric approach for simultaneously modeling multiple quantiles, and \cite{das2018bayesian} developed a non-parametric approach for complete and grid data. Recently, \cite{das2023bayesian} proposed a Bayesian hierarchical quantile regression that simultaneously models the hierarchical relationship of multi-level covariates.

In spite of the rich development on quantile regression approaches, the literature on models involving high-dimensional functional covariates is somewhat limited. \cite{cardot2005quantile} proposed a scalar-on-function quantile regression problem using a spline estimator to minimize a penalty term, while \cite{kato2012estimation} considered the principal component analysis method in a similar setting. \cite{zhang2022high} incorporated imaging data into the quantile regression as a functional response based on a function-on-scalar framework. \cite{wang2023bayesian} introduced the quantile latent-on-image regression with a two-stage model, using functional principal components of the images as predictors. Recent work has leveraged tensor-modeling approaches to tackle high dimensionality of spatially distributed imaging features under a quantile regression framework. 
\cite{lu2020high} first studied scalar on tensor quantile regression models based on Tucker decomposition and used regularization via $L_1$ penalties, while \cite{li2021tensor} proposed a tensor quantile regression model with a generalized LASSO penalty based on CANDECOMP/PARAFAC (CP) decompositions. \cite{wei2023tensor} proposed a Bayesian tensor response quantile regression to investigate the association between fMRI resting-state functional connectivity and PTSD symptom severity and \cite{liu2024tensor} developed a scalar-on-image tensor quantile regression based on tensor train (TT) decomposition with a generalized lasso penalty. In contrast to the above approaches that focused on cross-sectional imaging data, \cite{ke2023smoothed} proposed a two-stage estimation procedure for linear tensor quantile regression involving longitudinal tensor-valued data. They used generalized estimation equations for estimation, followed by kernel smoothing to construct smooth tensor coefficients in order to tackle high-dimensional covariates. As for Bayesian approaches,
\cite{reich2011bayesian} and \cite{reich2012spatiotemporal} considered spatiotemporal quantile regression models where each spatial location has its own quantile function that evolves over time and smoothed spatially via Gaussian process priors.  
\cite{wang2024nonparametric} developed a Bayesian nonparametric scalar-on-image quantile regression model based on Gaussian processes to analyze the association between cognitive decline and clinical variables. 

While the existing Bayesian approaches for quantile regression described above are attractive in providing uncertainty quantification and performing inference, they often rely on Gaussian process priors that may not be scalable to high-dimensional images containing tens of thousands of spatially distributed voxels. Moreover, approaches for Bayesian quantile regression approaches to map longitudinal neuroimaging changes are limited. 
In one of the first such attempts, we develop a novel Bayesian tensor quantile regression for high-dimensional longitudinal brain imaging data that uses tensor representations to model the longitudinal imaging coefficients in order to tackle high-dimensionality. Our goal is to estimate the longitudinal trajectory of brain-behavior associations and to infer significant neuroimaging biomarkers by pooling information across visits, while accurately predicting different quantiles of the outcome distribution. A key aspect of our proposed approach is the ability to differentiate between common associations across visits from visit-specific patterns contributing to longitudinal trajectories. This is made possible via a visit-invariant tensor coefficient that is learnt by systematically pooling information across visits, and additionally incorporating separate visit-specific tensor coefficients. 
Furthermore, a scalar random intercept term models the within-subject longitudinal dependencies in the outcome, while the similarity across visits is captured via the visit-invariant tensor coefficient term. We assume a low-rank PARAFAC decomposition on the tensor coefficients, preserving the spatial configuration of imaging voxels while addressing the challenges posed by the high dimensionality. 
We incorporate multiway shrinkage priors \citep{guhaniyogi2017bayesian} to model the visit-invariant tensor coefficients. We impose a spike and slab prior on the tensor margins for the visit-specific effects, which encourages sparsity and controls the degree of changes across visits from becoming exceedingly large.   We develop an MCMC algorithm for posterior inference that combines Gibbs sampling steps based on location-scale mixture representations with variable selection stochastic search algorithms.

 Our contributions are motivated by  ADNI-1 neuroimaging study, where the goal is to model quantiles of the cognitive score distribution based on T1-weighted MRI scans across longitudinal visits, and infer neuroimaging biomarkers driving cognitive deficits. Our analysis displays considerably improved prediction accuracy across various quantile levels consistently for both AD and MCI cohorts. Our results reveal sparse associations between the imaging voxels and the median cognitive scores, but more pronounced associations with tails of the distributions of the cognitive scores. 
Furthermore, our approach allows the identification of brain regions that are significantly associated with advanced cognitive impairment, which can serve as promising neuroimaging biomarkers. For instance, our analysis identified several gyral regions that showed significant associations with the 80-th quantile of the ADAS cognitive score (implying advanced cognitive impairment) but limited significant associations with the median or 20-th quantiles of the ADAS score distribution (indicating a more normal cognitive functioning). With respect to existing Bayesian quantile methods for functional covariates based on Gaussian processes, our analysis provides a more comprehensive understanding of disease progression and identification of neuroimaging biomarkers, by evaluating the longitudinal trajectory of neuroimaging effects on various quantiles of the cognitive score distribution. Additionally, we perform extensive simulation studies that show significant improvements in parameter estimation, feature selection, and prediction accuracy when compared to various competing methods.

The rest of the paper is organized as follows: In Section \ref{sec-method}, we describe the proposed tensor quantile regression model for longitudinal data and develop an MCMC algorithm for posterior inference. We also discuss multiplicity adjustments to control the Type 1 error. In Section \ref{sec-sim} we investigate the model performances through
comprehensive simulation studies. In Section \ref{sec-adni} we discuss our results from the data analysis using the ADNI dataset.

\section{Methods}\label{sec-method}

In its simplest form, a quantile regression model can be obtained from OLS regression as 
\begin{eqnarray}
Q_{q}(y_i) = \mathbf{x}_i^T \beta + \epsilon^q,  \mbox{ } i=1,\ldots,n, \label{eq:quant}
\end{eqnarray}
with $q$ denoting the quantile level of interest and $\epsilon^q$ the error term that is distributed such that it is zero at the $q$-th quantile. Our goal is to extend the classical Bayesian quantile regression framework in (\ref{eq:quant}) to the case of longitudinal studies involving high-dimensional spatial images as covariates, utilizing tensor-based methods to address the curse of dimensionality and incorporate spatial information in the model framework. 

\subsection{Bayesian Tensor Quantile Regression for Longitudinal Data}
We have available longitudinal MRI data on $n$ subjects, each of whom has a maximum of $T$ visits. The $i$th sample ($i=1,\ldots,n$) has $T_i \le T$ visits, where the visit index belongs to the set $\{1,2,\ldots,T \}$, with the understanding that subjects with missing visits will have data collected on a subset of visit indices $\{1,2,\ldots,T \}$. Let  $\mathcal{T}_{it}$ denote the time from baseline at the $t$-th visit for the $i$th individual, denote $\mathcal{X}_{it}$ as the brain image  for the $i$th subject at the $t$th visit, and denote the corresponding scalar outcome variable as $y_{it}$. Further, let ${\bf z}_i = (z_{i1},\ldots, z_{ip})$ denote the vector of time-invariant non-imaging covariates, for the $i$th subject. The proposed approach assumes regular visit times across samples, that is consistent with the design of ADNI study \citep{Jack2008ADNIMRIMethods} used for our data analysis. However, our approach can accommodate missing visits resulting in varying number of visits across samples, as elaborated in the sequel. 

We write our proposed Bayesian Longitudinal Tensor Quantile Regression (BLTQR) as:
\begin{eqnarray}
\label{model} 
Q_{q}(y_{i,t})= b_0 + b_{0i} + b_1*\mathcal{T}_{it} + {\bf z}^T_{i}  \mathbf{\eta} + \langle \mathcal{X}_{it}, \mathcal{B}_0 \rangle + \langle \mathcal{X}_{it}, \mathcal{B}_t \rangle + \epsilon^{q}_{it},
\end{eqnarray}
where $q$ is the quantile level, $b_0$ and $b_1$ represent the group-level intercept and slope, and $b_{0i}$ is the subject-specific intercept contributing to the variations in the longitudinal patterns across subjects. The effect of the brain image $\mathcal{X}_{it}$ is expressed as the sum of a visit-invariant effect ($\mathcal{B}_0$) and a visit-specific effect ($\mathcal{B}_t$), with both $\mathcal{B}_0$ and $\mathcal{B}_t$ tensors of dimensions $p_1\times p_2$, and with $\langle \cdot, \cdot \rangle$ in model \eqref{model} denoting the tensor inner product. The longitudinal trajectories are accounted for via visit-specific images $\mathcal{X}_{it}$, as well as the slope term. While the model can be easily generalized to include additional subject-specific random slope terms, we found the model (\ref{model}) to work sufficiently well in practical applications including the ADNI analysis in Section 4. In addition, we note that the model is naturally designed to handle missing visits. In particular,  the proposed model involves two types of coefficients: (a) a time-invariant parameter $\mathcal{B}_0$ and other scalar parameters that do not depend on the visit index and are estimated by combining information across all visits and samples; and (b) time-specific parameters $\mathcal{B}_t$ that is estimated by pooling information across all samples with data at the $t$-th visit, which is comparable across samples under the assumption of regular visit timings. For both (a) and (b) missing visits for a subset of samples do not alter the model implementation in any way. 

We complete our quantile regression model by assuming an asymmetric Laplace distribution (ALD) on the error term in equation \eqref{model} as
\begin{eqnarray*}
f(\epsilon^{q}_{it} \mid \sigma, q)=\frac{q(1-q)}{\sigma} \exp \left\{-\rho\left(\frac{\epsilon^{q}_{it}}{\sigma}\right)\right\}.
\end{eqnarray*}
\cite{kozumi2011gibbs} showed that ALD can be represented as a location-scale mixture of normal distributions where the mixing distribution follows an exponential distribution. Thus, equation \eqref{model} can be rewritten as 
\begin{eqnarray}
\label{model-ald} 
Q_{q}(y_{i,t})= b_0 + b_{0i} + b_1*\mathcal{T}_{it} +  {\bf z}^T_{i}  \mathbf{\eta} + \langle \mathcal{X}_{it}, \mathcal{B}_0 \rangle + \langle \mathcal{X}_{it}, \mathcal{B}_t \rangle\\ 
+ \theta \nu_{it}+ \rho\sqrt{\sigma \nu_{it}}u_{it},\nonumber
\end{eqnarray}
\noindent with $\nu_{it} = \sigma m_{it}$, $\theta = \frac{1-2q}{q(1-q)}$, and $\rho^2 = \frac{2}{q(1-q)}$, and where $m_{it}$ follows a standard exponential distribution and $u_{it}$ a standard normal distribution.

Our proposed model employs a tensor-based representation of the imaging data. 
A tensor is a multi-dimensional array, with the order being the number of its dimensions. For example, a one-way or first-order tensor is a vector, and a second-order tensor is a matrix. Mathematically, a \textit{N}-dimensional tensor $\mathcal{X}$ can be expressed as $\mathcal{X} \in \mathbb{R}^{I_1 \times I_2 \times \cdots \times I_N}$, where $I_n$ indicates the size of the tensor in the $n$-th dimension. Tensors have gained recognition as a modeling tool for neuroimaging data as they can achieve dimension reduction via low rank decompositions on the model parameters, while preserving the spatial information in the image.

Here we employ tensor decompositions for the visit-invariant effect ($\mathcal{B}_0$) and the visit-specific tensor effects ($\mathcal{B}_t$).
Tensor decomposition is a mathematical technique that breaks down a high-dimensional tensor into a combination of lower-dimensional factors. One type of tensor decomposition is the Tucker decomposition \citep{kolda2009tensor}, which decomposes a tensor into a core tensor and a set of matrices, one along each mode. 
Another commonly used technique is the PARAFAC decomposition, a special case of the Tucker decomposition, where the core tensor $\Lambda$ is restricted to be diagonal. The rank-\textit{R} PARAFAC model expresses the high-dimensional tensor as an outer produce of one-dimensional tensor margins, summed over $R$ channels.

Here, we employ PARAFAC decompositions that expresses the visit-invariant effect ($\mathcal{B}_0$) and the visit-specific tensor effects ($\mathcal{B}_t$) as  
\begin{eqnarray}
\mathcal{B}_0 = \sum_{r=1}^R \boldsymbol{\beta}_{01}^{(r)} \circ \cdots \circ \boldsymbol{\beta}_{0D}^{(r)},
\mbox{~and~} \mathcal{B}_t=\sum_{r_t=1}^{R_t} \boldsymbol{\chi}_{t1}^{(r_t)} \circ \cdots \circ \boldsymbol{\chi}_{tD}^{(r_t)}. 
\end{eqnarray}
\noindent where $\boldsymbol{\beta}_{01},\ldots, \boldsymbol{\beta}_{0D}$ and $\boldsymbol{\chi}_{t1},\ldots, \boldsymbol{\chi}_{tD}$, known as tensor margins, are vectors of length $p_1,\ldots,p_D$, and where $\boldsymbol{\beta}_{01} \circ \cdots \circ \boldsymbol{\beta}_{0D}$ and $\boldsymbol{\chi}_{t1} \circ \cdots \circ \boldsymbol{\chi}_{tD}$ are a D-way outer product of dimension $p_1 \times p_2 \times\ldots\times p_D$. For our neuroimaging applications, $D=2$ for 2-dimensional slices and $D=3$ for three-dimensional images. The assumed rank of the visit-specific effect ($R_t$) is allowed to be different from the assumed rank of the visit-invariant effect ($R$) to ensure that both effects are being modeled and optimized separately. The appropriate tensor rank vary depending on the specific application context and can be chosen using a goodness-of-fit approach \citep{guhaniyogi2017bayesian,liu2023integrative}. We note that, even though the tensor margins can only be uniquely identified up to a permutation and a multiplicative constant, the tensor product is fully identifiable, which suffices for our primary objective of estimating coefficients.  The PARAFAC decomposition dramatically reduces the number of coefficients from $p_1 \times \ldots \times p_D$ to $R(p_1 + \ldots + p_D)$, which grows linearly with the tensor rank $R$ and results in significant dimension reduction. 

Prior to applying the tensor model, the image's voxels are mapped onto a grid, making them more suitable for a tensor-based approach. This mapping conserves the spatial arrangements of the voxels. While the grid mapping might not preserve exact spatial distances between voxels, this has limited impact in our experience, since it can still capture correlations between neighboring elements in the tensor margins. Moreover, the tensor construction has the advantageous ability to estimate voxel-specific coefficients by leveraging information from neighboring voxels through the estimation of tensor margins with their inherent low-rank structure. This feature results in a type of spatial smoothing and results in brain maps that are more consistent and robust to missing voxels and image noise.  For more details on the characteristics of Bayesian tensor models, we refer readers to \cite{kundu2023bayesian}.

\subsection{Prior model}
Let's start with the prior choice for the tensor coefficients. For the tensor margins of the visit-unspecific tensor $\mathcal{B}_0$, we adopt the multiway Dirichlet generalized double Pareto (M-DGDP) prior from \cite{guhaniyogi2017bayesian}, which shrinks small coefficients towards zero while minimizing shrinkage of large coefficients. Let $\boldsymbol{W}_{j r}= \mbox{Diag}(w_{j r, 1}, \ldots, w_{j r, p_j})$, with $w_{jr,k}$, for $k=1,\ldots,p_j$, element-wise margin-specific scale parameters. The M-DGDP prior can be expressed in hierarchical form as:
\begin{align}
\label{GDPprior}
    \boldsymbol{\beta}_{0j}^{(r)} \sim & \mathrm{N}\left(0,\left(\phi_r \tau\right) \boldsymbol{W}_{j r}\right),\nonumber\\
        w_{j r, k}  \sim  \operatorname{Exp}&\left(\lambda_{j r}^2 / 2\right), \;\; \lambda_{j r} \sim \operatorname{Ga}\left(a_\lambda, b_\lambda\right),\nonumber\nonumber\\
        \left(\phi_1, \ldots, \phi_R\right) \sim & Dirichlet\left(\alpha_1, \ldots, \alpha_R\right),\\
    \tau \sim & Ga(a_{\tau}, b_{\tau}).\nonumber
    \end{align}
The parameter $\tau$ is a global scale parameter, while parameters $\Phi=\left(\phi_1, \ldots, \phi_R\right)$ encourage shrinkage to lower ranks in the assumed PARAFAC decomposition.
Marginalizing out the scale parameters $w_{j r, k}$,  one obtain 
\begin{equation}
\beta_{j, k}^{(r)} \mid \lambda_{j r}, \phi_r, \tau \stackrel{\mathrm{iid}}{\sim} \mathrm{DE}\left(\lambda_{j r} / \sqrt{\phi_r \tau}\right), 1 \leq k \leq p_j,
\end{equation}
that is, prior \eqref{GDPprior} induces a GDP prior on the individual margin coefficients which in turn has the form of an adaptive Lasso penalty \citep{armagan2013generalized}.

For the visit-specific effects $\mathcal{B}_t$, we adapt a form of the spike-and-slab prior to our setting, as we expect these signals to be sparse. Various spike-and-slab priors have been proposed in the field of Bayesian statistics, see \cite{vannucci2021} for an overview on the different approaches. These employ hierarchical prior structures and stochastic search MCMC methods to identify models with a high posterior probability of occurrence.  As the number of covariates increases, the task of determine the choice of priors becomes more challenging, since in many circumstances the high-frequency model is not the optimal predictive model \citep{barbieri2004optimal}. Here we adopt the prior construction of \cite{ishwaran2003detecting, ishwaran2005spike}, who employed a modified rescaled spike and slab models using continuous bimodal priors for the variance hyperparameters, and which has proved to be suitable for scenarios with large number of covariates. \cite{rovckova2014emvs} proposed a similar formulation using a discrete spike-and-slab model for the covariance parameters and employing the EM algorithm to rapidly identifies promising submodels. Let $\boldsymbol{W}^{\chi}_{tjr_t} = diag(w^{\chi}_{tjr_t,1},\ldots,w^{\chi}_{tjr_t,p_j})$. We impose spike-and-slab priors on the variance terms of the tensor margins of $\mathcal{B}_t$ as

\begin{equation} \label{ss-prior}
\begin{split}
    \boldsymbol{\chi}_{tj}^{(r_t)} &\sim N(0, (\phi_{tr_t}^{\chi}\tau_{t}^{\chi})\boldsymbol{W}^{\chi}_{tjr_t}) \\
     w^{\chi}_{tjr_t,k}|\pi_{tjr_t,k}, \; \lambda_{tjr_t}  &\sim (1-\pi_{tjr_t,k})\delta_{\nu} + \pi_{tjr_t,k}Exp(\lambda^{2}_{tjr_t})  \\
    \pi_{tjr_t,k} \sim Bern(\zeta_{tjr_t})&, \;\; \zeta_{tjr_t} \sim Beta(a_{\zeta}, b_{\zeta}), \\
      \left(\phi_{tr_t}, \ldots, \phi_{tR_t}\right) &\sim  Dirichlet\left(\alpha_1, \ldots, \alpha_R\right),\\
    \tau_{t}^{\chi} \sim  Ga(a_{\tau}, b_{\tau})&, \;\; \lambda_{tjr_t} \sim Ga(a_{\lambda}, b_{\lambda}).
\end{split}
\end{equation}
with $\delta_\nu$ denoting a point mass at $\nu$. In our construction, $\pi_{tjr_t,k}=1$ corresponds to large variances, encouraging large $ w_{tjr_t,k}$ values. With respect to the more commonly used spike-and-slab priors on individual regression coefficients, this construction has the computational advantage of retaining the block update for the tensor margins $\boldsymbol{\chi}_{tj}^{(r_t)}$, which drastically optimizes the computational cost of our MCMC sampling algorithm, as described in section \ref{mcmc}. To preserve the positive semi-definiteness of the covariance matrix, we set the spike at a small positive value that is nonzero ($\nu=10^{-4}$ in the applications of this paper). Finally, we complete the prior specification by assuming an Inverse Gamma distribution on $\sigma^2$, and standard Gaussian priors on $b_0$, $b_1$, $b_{0i}$, and $\mathbf{\eta}$.

\subsection{MCMC Algorithm}\label{mcmc}
For posterior inference, we employ a Markov Chain Monte Carlo (MCMC) algorithm that combines Gibbs sampling steps based on location-scale mixture representation of the asymmetric Laplace distribution \citep{kozumi2011gibbs}, and variable selection stochastic search algorithms that use add-delete-swap moves \citep{savitsky2011variable}. The algorithm is summarized below and the full details of the updates are provided in the Supplementary Material.

A generic iteration of the MCMC algorithm comprises of the following updates:
\begin{itemize}

\item Draw $\sigma$ from the full conditional

$$p(\sigma | \boldsymbol{y}, \mathcal{B}_0, \mathcal{B}_t, \nu_{it}, b_0, b_{0i}, b_1, \mathbf{\eta}) \sim \textrm{IG}(a/2, b/2),$$
with $a=n_0+3nT$ and $b=\frac{1}{\rho^2}\sum_{t=1}^T \sum_{i=1}^n \frac{(y_{it}-\mu_{it})^2}{\nu_{it}}+s_0+2\sum_{t=1}^T \sum_{i=1}^n \nu_{it}$, and where 
$$\mu_{it}=b_0 + b_{0i} + b_1*\mathcal{T}_{it}  + {\bf z}^T_{i}  \mathbf{\eta} + \langle \mathcal{X}_{it}, \mathcal{B}_0 \rangle + \langle \mathcal{X}_{it}, \mathcal{B}_t \rangle + \theta \nu_{it}.$$

\item Draw $\nu_{it}$ from the full conditional
$$p(\nu_{it} | y_{it}, \mathcal{B}_0, \mathcal{B}_t, \sigma, b_0, b_{0i}, b_1, \mathbf{\eta}) \sim \textrm{GIG}(\frac12, \, \delta_{it}, \, \gamma_{it}),$$
with $\delta_{it}^2=\frac{\tilde{y}_{it}^2}{\rho^2 \sigma}, \, \gamma_{it}^2 = \frac{\theta^2}{\rho^2 \sigma}+\frac{2}{\sigma}$, and $\tilde{y}_{it} = y_{it}-b_0 - b_{0i} - b_1*\mathcal{T}_{it}  - {\bf z}^T_{i}  \mathbf{\eta} - \langle \mathcal{X}_{it}, \mathcal{B}_0 \rangle - \langle \mathcal{X}_{it}, \mathcal{B}_t \rangle$.\\

\item Draw $b_0$ from the full conditional
$$p(b_0 | \boldsymbol{y}, \mathcal{B}_0, \mathcal{B}_t, \sigma,\boldsymbol{\nu}, b_{0i}, b_1, \mathbf{\eta}) \sim N(\mu_{b_0}, \sigma^2_{b_0}),$$ with $\sigma^2_{b_0}=(\sum^T_{t=1}\sum^n_{i=1}\frac1{\rho^2\sigma\nu_{it}})^{-1},\;\; \mu_{b_0}=\sigma^2_{b_0}(\sum^T_{t=1}\sum^n_{i=1}\frac{\tilde{y}_{it}}{\rho^2\sigma\nu_{it}})$, and $\tilde{y}_{it}=y_{it}-b_{0i}-b_1*\mathcal{T}_{it} -{\bf z}^T_{i}\mathbf{\eta}-\langle \mathcal{X}_{it}, \mathcal{B}_0 \rangle-\langle \mathcal{X}_{it}, \mathcal{B}_t \rangle-\theta \nu_{it}$.\\

\item Draw $b_{0i}$ from the full conditional
$$p(b_{0i} | \boldsymbol{y}_i, \mathcal{B}_0, \mathcal{B}_t, \sigma,\nu_{it}, b_{0}, b_1, \mathbf{\eta}) \sim N(\mu_{0i},\sigma^2_{0i}),$$ with $\sigma^2_{0i}=(\sum^T_{t=1}\frac1{\rho^2\sigma\nu_{it}}+1)^{-1};\;\; \mu_{0i}=\sigma^2_{0i}\sum^T_{t=1}\frac{\tilde{y}_{it}}{\rho^2\sigma\nu_{it}}$, and $\tilde{y}_{it}=y_{it}-b_0-b_1*\mathcal{T}_{it}-{\bf z}^T_{i}\mathbf{\eta}-\langle \mathcal{X}_{it}, \mathcal{B}_0 \rangle-\langle \mathcal{X}_{it}, \mathcal{B}_t \rangle-\theta \nu_{it}$.\\

\item Draw $b_1$ from the full conditional 
$$p(b_1 | \boldsymbol{y}, \mathcal{B}_0, \mathcal{B}_t, \sigma,\nu_{it}, b_{0}, b_{0i}, \mathbf{\eta}) = N(\mu_{b_1}, \sigma^2_{b_1}),$$ with $\sigma^2_{b_1}=(\sum^T_{t=1}\sum^n_{i=1}\frac{\mathcal{T}_{it}^2}{\rho^2\sigma\nu_{it}}+1)^{-1};\;\; \mu_{b_1}=\sigma^2_{b_1}(\sum^T_{t=1}\sum^n_{i=1}\frac{\tilde{y}_{it}\mathcal{T}_{it}}{\rho^2\sigma\nu_{it}})$, and $\tilde{y}_{it}=y_{it}-b_0-b_{0i} -{\bf z}^T_{i}\mathbf{\eta}-\langle \mathcal{X}_{it}, \mathcal{B}_0 \rangle-\langle \mathcal{X}_{it}, \mathcal{B}_t \rangle-\theta \nu_{it}$.\\

\item Update $[\alpha, \Phi, \tau \mid \mathcal{B}_0, \boldsymbol{W}]$ compositionally as $[\alpha \mid \mathcal{B}_0, \boldsymbol{W}][\Phi, \tau \mid \alpha, \mathcal{B}_0, \boldsymbol{W}]$, as described in \cite{guhaniyogi2017bayesian}.\\

\item Sample $\{(\beta_{0j}^{(r)}, w_{jr}, \lambda_{jr}); \; 1\leq j \leq D,\; 1 \leq r \leq R\}$ using a back-fitting procedure to produce a sequence of draws from the margin-level conditional distributions across components. For $r = 1,...,R$ and $j = 1,...,D$, sample from the conditional distribution $\left[\left(\boldsymbol{\beta}_{0j}^{(r)}, w_{j r}, \lambda_{j r}\right) \mid \boldsymbol{\beta}_{0,-j}^{(r)}, \mathcal{B}_{0,-r}, \boldsymbol{y}, \Phi, \tau, \mathbf{\eta},  \sigma, \boldsymbol{\nu}\right]$, where $\boldsymbol{\beta}_{0,-j}^{(r)}=\left\{\boldsymbol{\beta}_{0l}^{(r)}, l \neq j\right\}$ and $\mathcal{B}_{0,-r}=\mathcal{B} \backslash \mathcal{B}_{0r}$, as:

\begin{itemize}
    \item[(a)] Draw $\left[w_{j r}, \lambda_{j r} \mid \boldsymbol{\beta}_{0j}^{(r)}, \phi_r, \tau\right]=\left[w_{j r} \mid \lambda_{j r}, \boldsymbol{\beta}_{0j}^{(r)}, \phi_r, \tau\right]\left[\lambda_{j r} \mid \boldsymbol{\beta}_{0j}^{(r)}, \phi_r, \tau\right]$.

    \begin{enumerate}
        \item Draw $\lambda_{jr} \sim \mathrm{Ga}\left(a_\lambda+p_j, b_\lambda+\left\|\boldsymbol{\beta}_{0j}^{(r)}\right\|_1 / \sqrt{\phi_r \tau}\right)$;
        \item Draw $w_{jr,k} \sim \operatorname{GIG}\left(\frac{1}{2}, \lambda_{j r}^2, \beta_{0j, k}^{2(r)} /\left(\phi_r \tau\right)\right)$ independently for $1 \leq k \leq p_j$.
    \end{enumerate}

    \item[(b)] Draw $\beta_{0j}^{(r)} \sim N(\mu_{0jr}^{\beta}, \Sigma_{0jr}^{\beta})$. See full expressions of $\mu_{0jr}^{\beta}$ and $\Sigma_{0jr}^{\beta})$ in the Supplementary Material.\\
\end{itemize}

\item Update $[\alpha_t^{\chi}, \Phi_{tr_t}^{\chi}, \tau_t^{\chi} \mid \mathcal{B}_t, \boldsymbol{W}^{\chi}_{tjr_t}]$ compositionally as $[\alpha_t^{\chi} \mid \mathcal{B}_t, \boldsymbol{W}^{\chi}_{tjr_t}][\Phi_{tr_t}^{\chi}, \tau_t^{\chi} \mid \alpha_t^{\chi}, \mathcal{B}_t, \boldsymbol{W}^{\chi}_{tjr_t}]$, as described in \cite{guhaniyogi2017bayesian}.\\

\item Jointly update $(\pi^{(r_t)}_{tj},\;w^{\chi}_{tjr_t})$. This is a joint Metropolis-Hastings step. We use the add/delete/swap method to propose a new candidate $\pi^{*}_{tjr_t,k}$, and update the added/deleted or swapped voxels $w^{\chi}_{tjr_t})$ based on the proposed $\pi^{*}_{tjr_t,k}$. The proposed candidate is jointly accepted or rejected. See Supplementary Material for full details of the acceptance probability.\\

\item Sample $\{(\boldsymbol{\chi}_{tj}^{(r_t)}, \lambda_{tjr_t}); \; 1\leq j \leq D,\; 1 \leq r_t \leq R_t\}$ using a back-fitting procedure to produce a sequence of draws from the margin-level conditional distributions across components. For $r_t = 1,...,R_t$ and $j = 1,...,D$, sample from the conditional distribution $\left[\left(\boldsymbol{\chi}_{tj}^{(r_t)},  \lambda_{tjr_t})\right) \mid \boldsymbol{\chi}_{t,-j}^{(r_t)}, \mathcal{B}_{t,-r_t}, \boldsymbol{y}_t, \Phi_t^{\chi}, \tau_t^{\chi}, \mathbf{\eta},  \sigma, \boldsymbol{\nu}\right]$. as:

\begin{itemize}
    
    \item Draw $\lambda_{tjr_t} \sim \mathrm{Ga}\left(a_\lambda+p_j, b_\lambda+\left\|\boldsymbol{\chi}_{tj}^{(r)}\right\|_1 / \sqrt{\phi^{\chi}_{t,r_t} \tau^{\chi}_t}\right)$;

    \item Draw $\boldsymbol{\chi}_{tj}^{(r_t)} \sim N(\mu_{tjr_t}^{\chi}, \Sigma_{tjr_t}^{\chi})$. See full expressions of $\mu_{tjr_t}^{\chi}$, and $\Sigma_{tjr_t}^{\chi}$ in the Supplementary Material.\\
\end{itemize}

\item Draw $\zeta_{tjr_t}$ from 
$$p(\zeta_{tjr_t} | \pi_{tjr_t}) \sim Beta(\alpha,\beta),$$
with $\alpha = a_{\zeta}+\sum_{k=1}^{p_j}(\pi^{(r_t)}_{tj,k}$ and $\beta = b_{\zeta}+\sum_{k=1}^{p_j}(1-\pi^{(r_t)}_{tj,k}$.\\

\item Draw $\mathbf{\eta}$ from 
$$p(\mathbf{\eta} | \boldsymbol{y}, \mathcal{B}_0, \mathcal{B}_t, \sigma, \boldsymbol{\nu}, b_0, b_{0i}, b_1) \sim N(\mu_{\eta}, \Sigma_{\eta}),$$
with $\Sigma_{\eta}^{-1}=\sum_{t=1}^T \frac{{\bf z}^T\boldsymbol{1/\nu_{t}} {\bf z}}{\rho^2 \sigma} + \Sigma_{0\eta}^{-1}, \, \mu_{\eta} = \Sigma_{\eta}(\sum_{t=1}^T \frac{{\bf z}^T \tilde{\bf y}}{\rho^2 \sigma \boldsymbol{\nu}_t})$, and $\tilde{y}_{it}=y_{it}-b_0 - b_{0i} - b_1*\mathcal{T}_{it} - \langle \mathcal{X}_{it}, \mathcal{B}_{0l} \rangle - \langle \mathcal{X}_{it}, \mathcal{B}_{t} \rangle - \theta \nu_{it}$.

\end{itemize}

\subsection{Posterior Inference}
\label{multiplicity}
Given the MCMC output, we obtain estimates of the cell-level tensor effects for each longitudinal visit and achieve feature selection as follows. 

In our model formulation, the tensor effect per visit is captured by both the estimated visit-specific and visit-unspecific effects as $(\hat{\mathcal{B}_0} + \hat{\mathcal{B}_t})$. 
Let $\theta_{tj},~j = 1,\ldots, J$ be the vectorized tensor coefficients at visit $t$, with $J=\prod_{k=1}^D p_k$ the total number of cells in the image $\mathcal{X}_t$. Suppose $K$ posterior samples across $J$ elements of a tensor-valued coefficient denoted as $\left\{\Gamma^k=\left(\Gamma^k\left(\theta_1\right), \ldots, \Gamma^k\left(\theta_J\right)\right)^{\prime}: k=1, \ldots, K\right\}$ were obtained after burn-in. We obtain the significant coefficient estimates with multiplicity adjustments. Multiplicity adjustment is a widely-used technique to control the increased risk of Type I errors, i.e., the false discovery rate. Specifically, we used the “Mdev” method, which relies on simultaneous credible bands, introduced in \citep{crainiceanu2007spatially} and later adopted by \cite{hua2015semiparametric} and \cite{kundu2023bayesian}. Given the posterior sample curve $\widehat{\Gamma\left(\theta_j\right)}$ and the point-wise $\alpha/2$ and $1-\alpha/2$ quantiles $\zeta_{\alpha/2}\left(\theta_j\right)$ and $\zeta_{1-\alpha/2}\left(\theta_j\right)$, the Mdev method computes the maximal deviations $\zeta_{\alpha/2}^*=\max _{j=1, \ldots, J} \left(\widehat{\Gamma\left(\theta_j\right)} - \zeta_{\alpha/2}\left(\theta_j\right)\right)$ and $\zeta_{1-\alpha/2}^*=\max _{j=1, \ldots, J} \left(\zeta_{1-\alpha/2}\left(\theta_j\right) - \widehat{\Gamma\left(\theta_j\right)}\right)$. The credible bands for each voxel are obtained as $\left[\widehat{\Gamma\left(\theta_j\right)}-\zeta_{\alpha/2}^*, ~\widehat{\Gamma\left(\theta_j\right)}+\zeta_{1-\alpha/2}^*\right]$, where the voxel $\theta_j$ is considered significant if the corresponding credible interval does not contain zero. 

\section{Simulation Study}\label{sec-sim}
In this section, we assess model performance via several distinct simulated scenarios, where we consider both visit-varying and visit-invariant effects, and compare results with competing methods.

\subsection{Data Generation}

We generate the data under several scenarios. To assess the proposed method's ability to retain spatial information of the images, while pooling information across spatially distributed voxels and longitudinal visits, we generate data from 3 visits with the varying signal shapes. For each visit, the signal image is generated with dimension 48 by 48 for Scenarios 1-4, and from a $30\times 30\times 30$ 3-D image in Scenario 5. For each scenario, we varied the size and magnitudes of these signals while keeping the signal shape constant across visits. In addition, we included some sparse signals with magnitudes ranging from 1.2 to 2 at the last visit for all scenarios. Notably, the coefficient signal shapes used for data generation did not follow the low rank PARAFAC decomposition that is implicit under our modeling construction.

\begin{itemize}
    \item {\bf Scenario 1}: Rectangular signals varying in locations and magnitudes, with additional sparse signals in the last visit.
    \item {\bf Scenario 2}: Cross-shaped signals varying in locations and magnitudes, with additional sparse signals in the last visit.
    \item {\bf Scenario 3}: Triangular signals varying in locations and magnitudes, with additional sparse signals in the last visit.
    \item {\bf Scenario 4}: Circular signals varying in locations and magnitudes, with additional sparse signals in the last visit.
    \item {\bf Scenario 5}: Cubic 3-D shapes for $30\times 30\times 30$ images varying in locations and magnitudes, with additional sparse signals at the last visit.
\end{itemize}

\begin{figure}[h]%
\centering
\includegraphics[width=1\textwidth]{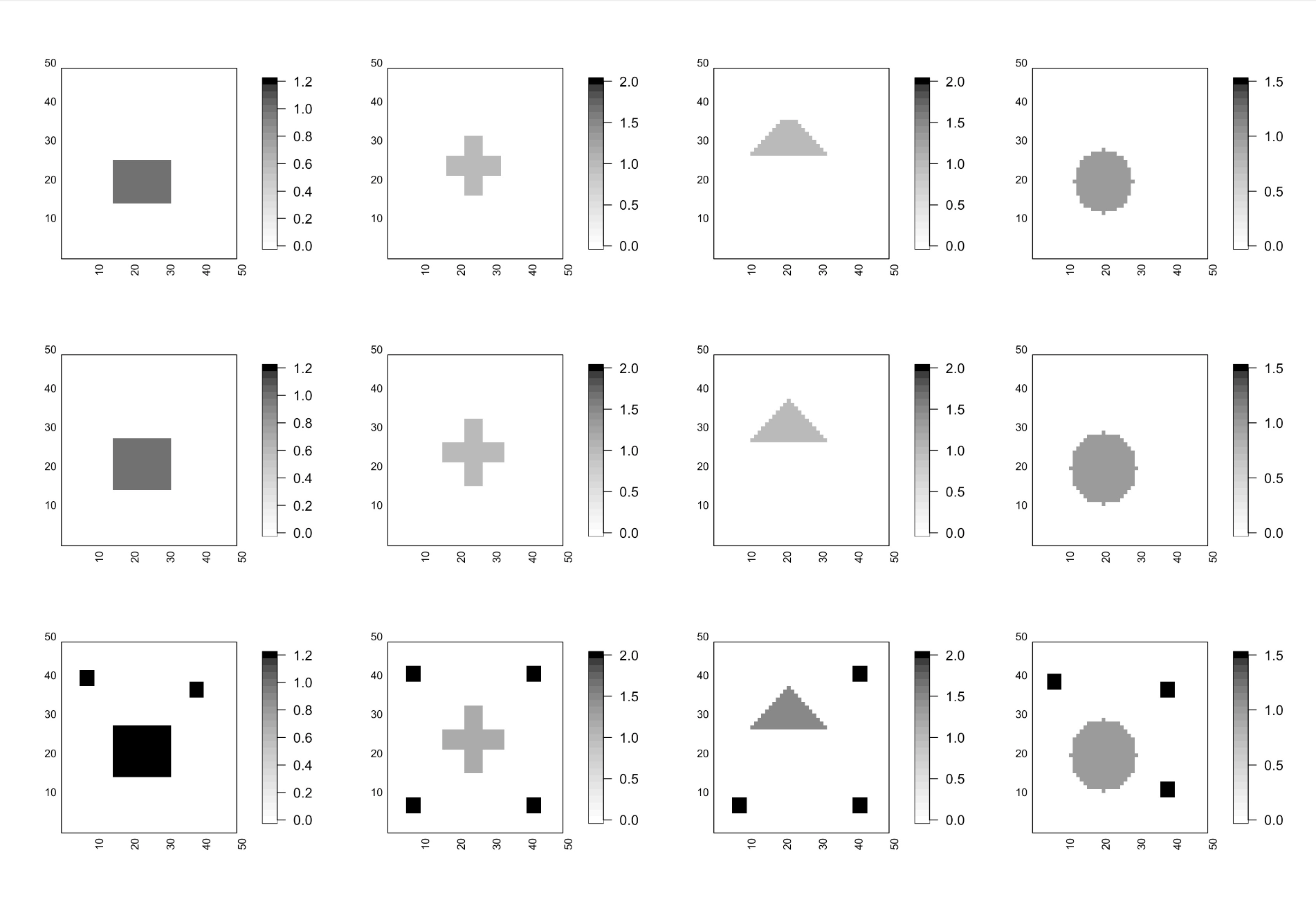}
\caption{Simulated signals for scenarios 1-4, as described in the text. Each column represents one scenario, with rows 1-3 representing the longitudinal visits.}
\label{Sim_signals}
\end{figure}

Figure \ref{Sim_signals} shows the true longitudinal 2D tensor signals for four scenarios. For each setting, the tensor covariates $\mathcal{X}_t$ were generated from the standard normal distribution $N(0,1)$, and the random error $\epsilon_t$ was generated from $ALD(0,\sigma = 1, q)$, with the quantile level $q$ chosen to be 0.2, 0.5, and 0.8. For each simulation, we generate 250 samples per visit for training, and addition 50 test samples were generated for evaluating out-of-sample prediction.

\begin{figure}[h]%
\centering
\includegraphics[width=1\textwidth]{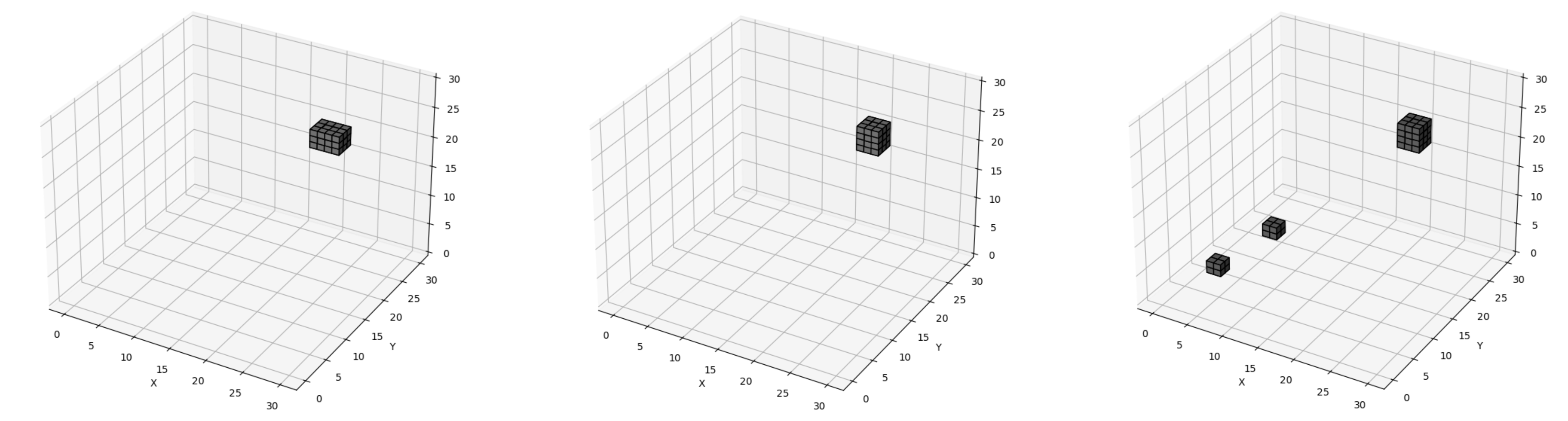}
\caption{Simulated signals for 3D setup. From left to right: visit 1( with magnitude 1), visit 2( with magnitude 1), and visit 3( with magnitude 1.5).  }
\label{3D_sim_signals}
\end{figure}

\subsection{Parameter Settings}
We use the default values suggested by \cite{guhaniyogi2017bayesian} for the prior hyper-parameters of the Bayesian tensor regression model. Specifically, hyperparameter of the Dirichlet component $\left(\alpha_1, \ldots, \alpha_R \right) =\alpha =  (1/R, ..., 1/R)$, $a_\lambda = 3$,  $b_\lambda=\sqrt[2 D]{a_\lambda}$, $a_\tau = \sum_{i=1}^R \alpha_i$, and $b_\tau = \alpha R^{(1/D)} $ are set as defaulted values to control the cell-level variance on tensor $\boldsymbol{B}$. When choosing suitable values of the prior hyperparameters, for the multiway shrinkage prior we set the parameters of the hyperprior on the global scale $\tau$ to $a_\tau = 1$ and $b_\tau = \alpha R^{(1/D)}$, where $R$ is the rank in the assumed PARAFAC decomposition, and set $\alpha_1 = \ldots = \alpha_R = 1/R$. For the common rate parameter $\lambda_{jr}$, we set $a_{\lambda} = 3$ and $b_\lambda=\sqrt[2 D]{a_\lambda}$. For the spike-and-slab prior, we set the hyperparameters of the Bernoulli probability parameter $\delta_{tjr_t}$ to be $a=1$ and $b=1$. 

In order to decide the rank of the fitted model, we fit the proposed model using ranks 2-4 and choose the rank that minimized the Deviance Information Criterion (DIC) scores. DIC measures the goodness-of-fit while adjusting for model complexity. When fitting our model, we ran 7000 iterations for each MCMC chain with 2500 burnins.

\subsection{Performance Metrics}
Performance metrics were selected to assess estimation accuracy and feature selection performance of the cell-level tensor effects for each longitudinal visit and to evaluate the prediction accuracy of the quantile estimates.  We used the following metrics to assess point estimation performance: 
\begin{itemize}
\item[(i)] 
Relative error of $\theta_t$, denoted as $RE(\theta_t) = \frac{\sum_{j=1}^{J}|\hat{\theta}_{tj} - \theta_{tj}|}{\sum_{j=1}^{J}|\theta_{tj}|}$, that measures the scaled absolute deviations between the true and estimated parameters,  with $\hat{\theta}_{tj}$ and $\theta_{tj}$ the estimated and true coefficients, respectively. We note that the values of relative error can be larger than 1, since the discrepancy between estimated and true values of tensor cells can be greater than zero when a large portion of the true values is exactly 0. \item[(ii)] Root-mean-square error of $\theta_t$, denoted as $RMSE(\theta_t) = \sqrt{\frac{\sum_{j=1}^{J} (\hat{\theta}_{tj} - \theta_{tj})^2}{J}}$.
\item[(iii)] correlation coefficient between the true and estimated cell-level tensor coefficients.
\end{itemize}   

\noindent For feature selection, performances were evaluated by the following metrics:
\begin{itemize}
\item[(i)] Sensitivity = $\frac{TP}{TP+FN}$; 
\item[(ii)] Specificity = $\frac{TN}{TN+FP}$; 
\item[(iii)] MCC = $\frac{TP\times TN-FP \times FN}{\sqrt{(TP+FP)(TP+FN)(TN+FP)(TN+FN)}}$, 
\end{itemize}
with TP the true positive, i.e. the number of predictions where the classifier correctly predicts a positive class as positive; FP the false positive, i.e. the number of predictions where the classifier incorrectly predicts a negative class as positive; TN the true negative, i.e. the number of predictions where the classifier correctly predicts a negative class as negative; and FN the false negative, i.e. the number of predictions where the classifier incorrectly classifies a positive class as negative. In our scenario, the positive class refers to the non-zero voxels of the tensor coefficients, and the negative class refers to the zero voxels of the tensor coefficients.

Finally, prediction accuracy was evaluated based on the check loss function \citep{wang2012quantile} as $E(l(Y_{t}, \hat{Q}_q(Y_{t})))$, where
$$
l(Y_{it}, \hat{Q}_q(Y_{it}))= \begin{cases}q|Y_{it}-\hat{Q}_q(Y_{it})|, & Y_{it}>\hat{Q}_q(Y_{it}) \\ (1-q)|Y_{it}-\hat{Q}_q(Y_{it})|, & Y_{it} \leq \hat{Q}_q(Y_{it})\end{cases},
$$
and with $\hat{Q}_q(Y_{it})$ the predicted $q$-th quantile for subject $i$ at visit $t$.

\subsection{Results}\label{sim-results}
In this section, we present the point estimation, feature selection, and out-of-sample prediction performance for quantile level 0.5. The complete results for quantile levels 0.2 and 0.8 are available in Section 2 of the Supplementary Material.
Figure \ref{Sim_est_signals} shows the 2D tensor estimates obtained using the proposed method, on one replicate for each of the 4 scenarios. From this Figure, it is evident that the proposed method is able to broadly recover the shapes of the 2D tensor coefficient for each visit, and any sparse signals present in the last visit regardless of the complexity of the different shapes.

\begin{figure}[h]
\centering
\includegraphics[width=1\textwidth]{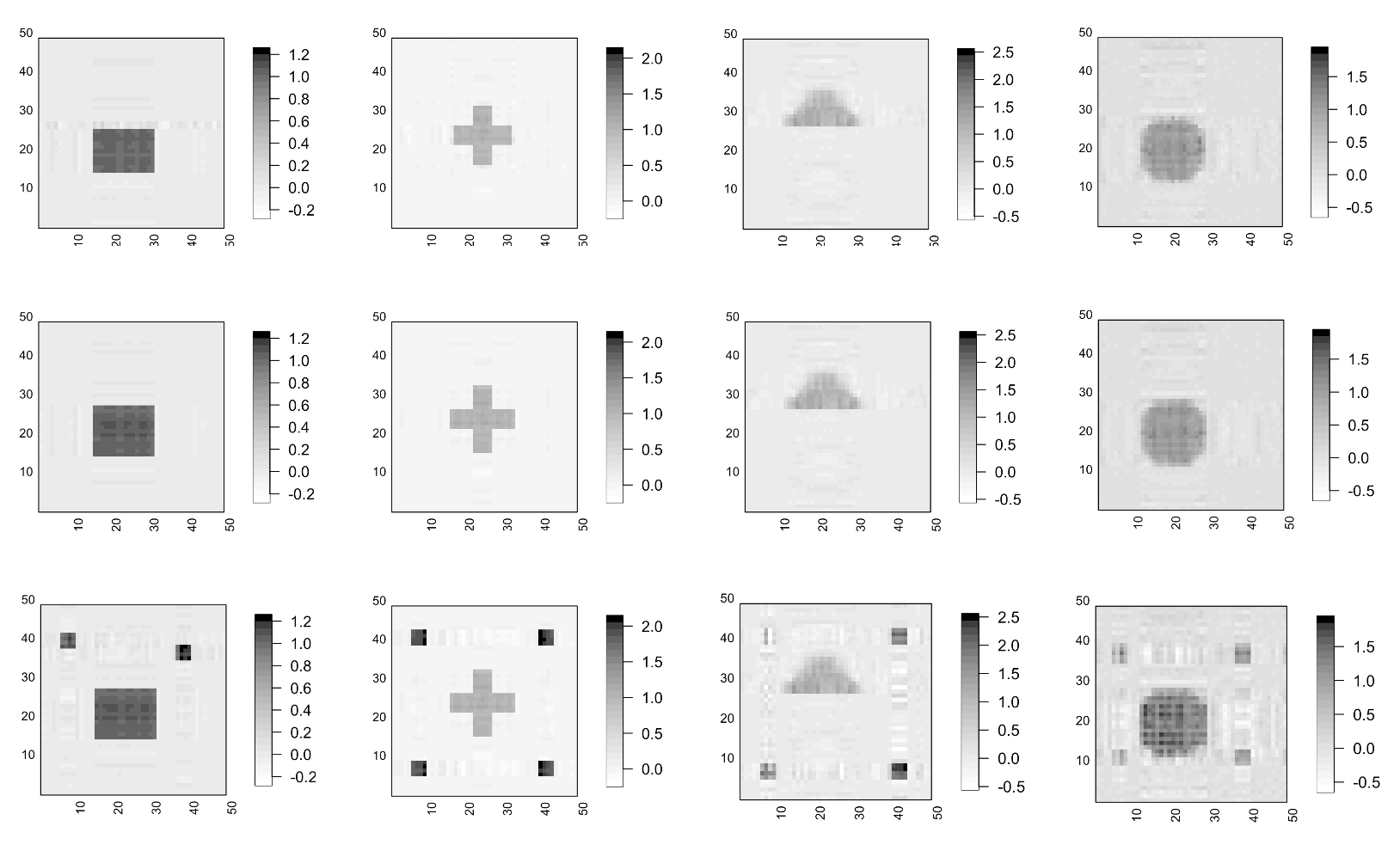}
\caption{Estimated signals for one replicate of scenarios 1-4, as described in the text. Each column represents one scenario, with rows 1-3 representing the longitudinal visits.}
\label{Sim_est_signals}
\end{figure}

Results under the proposed approach and competing methods for parameter estimation, feature selection, and prediction, averaged over 10 replicates, are reported in Tables \ref{sim-pe}, \ref{sim-fs}, \ref {sim-mae}, respectively. We consider four alternative methods: (i) Cross-sectional Bayesian tensor quantile regression type 1 (CsB1), obtained from our model by fixing  $\mathcal{B}_0$ to be 0 in equation \eqref{model}, and placing the M-DGDP prior as in equation \eqref{GDPprior} on $\mathcal{B}_t$; (ii) Cross-sectional Bayesian tensor quantile regression type 2 (CsB2), obtained from our model by fixing $\mathcal{B}_t$ to be 0 in equation \eqref{model}, and placing the M-DGDP prior as in equation \eqref{GDPprior} on $\mathcal{B}_0$; (iii) the tensor quantile regression (Freq. TQR) of \cite{li2021tensor} that performs frequentist cross-sectional tensor quantile regression, with the implementation made publicly available by the authors in MATLAB; and (iv) Quantile regression (rqPen) with the lasso penalty from R package \texttt{rqPen} that does not account for the spatial information in the images.
 CsB1 essentially performs a cross-sectional analysis by removing the shared parameters $\mathcal{B}_0$ and performing the analysis separately at each visit. In contrast, CsB2 performs another type of cross-sectional analysis by stacking data from all visits to fit a common model without accounting for visit-specific differences. The two other frequentist approaches considered also fit their model separately across different visits. The feature selection results were adjusted for multiplicity as described in Section \ref{mcmc}. It was not possible to report feature selection under the frequentist tensor quantile regression approach as this method uses the Black Relaxation Algorithm to solve optimization problems and does not inherently provide confidence intervals.

Results presented in Tables \ref{sim-pe}-\ref{sim-fs} demonstrate that the proposed model outperforms all alternative methods in both coefficient estimation and feature selection, across all scenarios. In particular, the proposed approach consistently has improved RE and RMSE compared to all competing approaches under all scenarios. The CsB1 approach is seen to have a comparable estimation compared to the proposed approach for visits 1 and 2 under Scenario 1, however the estimation is considerably poorer for other scenarios. For Scenario 5 including 3-D images, CsB 2 has improved point estimation for visit 1, while the point estimation is inferior relative to the proposed approach for other visits. Given that visit 3 has sparse signals, the results potentially suggest that the proposed approach is better suited to capture sparse signals by separating out the common effects across visits and imposing spike and slab priors for the visit-specific effects. In terms of feature selection (as reported in Table \ref{sim-fs}), the proposed approach consistently has higher or comparable F-1 and MCC scores compared to alternate methods across all scenarios when averaged across visits. While the CsB1 method sometimes has comparable F-1 and MCC scores for visits 1 and 2, its feature selection for visit 3 is considerably inferior across all settings. Moreover, the proposed approach shows consistently higher F-1 and MCC scores across all visits for Scenario 5 involving 3-D images. The superior performance under the proposed method for visit 3 demonstrates the advantages of imposing a selection prior on the visit-specific tensor coefficients for selecting significant cells. 

\begin{table*}[hbt!]
\caption{Point Estimation results of cell-level signals for the 5 scenarios portrayed in Figure \ref{Sim_signals} corresponding to $q=0.5$.}\label{sim-pe}
\begin{tabular*}{\textwidth}{@{\extracolsep\fill}ll|ccc|ccc|ccc}
\toprule%
 \multirow{2}{4em}{Scenarios} &  &  \multicolumn{3}{c}{RE} & \multicolumn{3}{c}{RMSE}& \multicolumn{3}{c}{Corr. Coef.}\\
\cline{3-5}
\cline{6-8}
\cline{9-11}
& & v1&v2&v3&v1&v2&v3&v1&v2&v3\\
\midrule
\multirow{5}{4em}{1} &BLTQR & \bf{0.159}& \bf{0.125}& \bf{0.254}&0.026&\bf{0.021}&\bf{0.058}&\bf{0.996}&\bf{0.998}&\bf{0.994}\\
& CsB1&0.165 &0.143 &0.370& \bf{0.022} &0.023 &0.081 &\bf{0.996} &0.997 &0.975 \\
& CsB2 &0.656 &0.493 &0.546& 0.101 &0.068 &0.160 &0.938 &0.971 &0.912\\
&Freq. TQR &0.221 &0.196 &0.875 &0.029 &0.030 &0.206 &0.994 &0.994&0.841 \\
&rqPen &1.080 &1.082 &1.026 &0.277 &0.306 &0.390 &0.165 &0.172&0.047 \\
\midrule
\multirow{5}{4em}{2} &BLTQR & \bf{0.291}& \bf{0.217}& \bf{0.258} &\bf{0.031} &\bf{0.024} &\bf{0.060} &\bf{0.991} &\bf{0.995} &\bf{0.992}\\
&CsB1 &0.405 &0.390 &1.313 &0.040 &0.044 &0.267 &0.985 &0.985&0.736\\
&CsB2 &1.604 &1.345 &0.971 &0.138 &0.128 &0.304 &0.852 &0.873 &0.729\\
&Freq. TQR &1.675 &1.713 &2.473 &0.161 &0.187 &0.512 &0.746 &0.683 &0.306\\
&rqPen &0.960 &1.045 &1.004 &0.224 &0.252 &0.450 &0.267 &0.105&0.004\\
\midrule
\multirow{5}{4em}{3}&BLTQR &\bf{0.701}& \bf{0.697}& \bf{1.037}& \bf{0.069} &\bf{0.075}& \bf{0.218}&\bf{0.951} &\bf{0.942}& \bf{0.868}\\
&CsB1 &0.946 &1.015 &2.121 &0.098 &0.107 &0.403 &0.896 &0.877 &0.498\\
&CsB2 &2.393 &2.327 &1.380 &0.173 &0.175 &0.325 &0.780 &0.774 &0.674\\
&Freq. TQR &1.571 &1.717 &2.821 &0.145 &0.163 &0.545 &0.775 &0.713 &0.214\\
&rqPen &1.061 &1.276 &0.997 &0.216 &0.227 &0.445 &0.295 &0.320 &0.128\\
\midrule
\multirow{5}{4em}{4}&BLTQR & \bf{0.524}& \bf{0.382}& \bf{0.447}&\bf{0.094} &\bf{0.087}& \bf{0.115}&\bf{0.945} &\bf{0.962}& \bf{0.952}\\
&CsB1 &0.638 &0.543 &0.946 &0.103 &0.110 &0.237 &0.929 &0.937 &0.816\\
&CsB2 &1.102 &0.838 &0.828 &0.150 &0.143 &0.241 &0.869 &0.892 &0.767\\
&Freq. TQR &0.785 &0.747 &1.244 &0.132 &0.149 &0.321 &0.888 &0.883&0.685\\
&rqPen &1.249 &1.192 &0.996 &0.305 &0.351 &0.393 &0.195 &0.167 &0.116\\
\midrule
\multirow{5}{4em}{5}&BLTQR & 1.180&{\bf 0.803}&{\bf 1.131}&0.009&0.007&{\bf 0.018}&0.968&{\bf 0.985} &{\bf 0.971}\\
& CsB1&6.165 &5.925 &3.189 &0.025 &0.017 &0.044 &0.799 &0.906 &0.756 \\
& CsB2 & {\bf 0.880} &0.997 &1.605 &{\bf 0.006} &0.007 &0.022 &{\bf 0.984} &0.978 &0.946\\
&Freq. TQR &11.638 &12.596 &11.925 & 0.043 &0.044 &0.091 &0.230 &0.188 &0.002 \\
&rqPen &1.119 &1.016 &1.245 &0.037 &0.036 &0.065 &0.021 &0.000 &0.147 \\
\bottomrule
\end{tabular*}
\end{table*}

\begin{table*}[hbt!]
{\footnotesize
\caption{Feature selection results for the 5 scenarios portrayed in Figure \ref{Sim_signals} corresponding to $q=0.5$.}\label{sim-fs}
\begin{tabular*}{\textwidth}{@{\extracolsep\fill}p{0.2cm}p{0.77cm}p{0.5cm}p{0.5cm}p{0.5cm}p{0.5cm}p{0.5cm}p{0.5cm}p{0.5cm}p{0.5cm}p{0.5cm}p{0.5cm}p{0.5cm}p{0.5cm}}
\toprule%
 \multirow{2}{4em}{Scenarios} & &  \multicolumn{3}{c}{Sens.} & \multicolumn{3}{c}{Spec.}& \multicolumn{3}{c}{F1-score}& \multicolumn{3}{c}{MCC}\\
\cline{3-5}
\cline{6-8}
\cline{9-11}
\cline{12-14}
& &v1&v2&v3&v1&v2&v3&v1&v2&v3&v1&v2&v3\\
\midrule
 \multirow{4}{4em}{1}&BLTQR &\bf{1}&\bf{1}&\bf{1}&\bf{1}&\bf{1}&\bf{1}&\bf{1}&\bf{1}&\bf{1}&\bf{1}&\bf{1}&\bf{1}\\
&CsB1& \bf1&\bf1 &0.977&\bf1&\bf1&\bf1&\bf1&\bf1&0.988 &\bf1&\bf1&0.987\\
&CsB2&\bf1&\bf1&0.88 &1& 0.997& 0.998& 0.904& 0.987& 0.931& 0.900& 0.986& 0.926\\
&rqPen &0.147 &0.158 &0.016& 0.970& 0.963& 0.990& 0.197& 0.208& 0.030& 0.164& 0.165& 0.021\\
\cline{2-5}
\cline{6-8}
\cline{9-11}
\cline{12-14}
 \multirow{4}{4em}{2}&BLTQR &\bf{1}&\bf{1}&\bf{1}&0.999&\bf{1}&\bf{1}&0.999&\bf{1}&\bf{1}&0.999&\bf{1}&\bf{1}\\
&CsB1& \bf1&\bf1 &0.333&\bf{1}&\bf1&0.999&\bf{1}&\bf1&0.422 &\bf{1}&\bf1&0.526\\
&CsB2&\bf1&\bf1&0.864 &0.960& 0.969& 0.985& 0.747& 0.818& 0.861& 0.757& 0.820& 0.849\\
&rqPen &0.120 &0.076 &0.005& 0.995& 0.985& 0.999& 0.199& 0.117& 0.009& 0.246& 0.109& 0.042\\
\cline{2-5}
\cline{6-8}
\cline{9-11}
\cline{12-14}
 \multirow{4}{4em}{3}&BLTQR &\bf0.997&\bf0.967&0.647&\bf0.997&\bf0.997&\bf0.999&\bf0.968&\bf0.956&\bf0.777&\bf0.967&\bf0.953&\bf0.788\\
&CsB1& 0.748&0.684 &0.108&0.999&0.999&0.999&0.847&0.802&0.172 &0.851&0.812&0.259\\
&CsB2&0.953&0.928&\bf{0.737} &0.969& 0.970& 0.975& 0.774& 0.768& 0.975& 0.778& 0.770& 0.708\\
&rqPen &0.256 &0.421 &0.030& 0.960& 0.915& 0.997& 0.256& 0.285& 0.055& 0.217& 0.246& 0.095\\
\cline{2-5}
\cline{6-8}
\cline{9-11}
\cline{12-14}
\multirow{4}{4em}{4}&BLTQR &0.878&0.876&0.758&\bf0.999&\bf0.999&\bf1& \bf 0.917&\bf0.943&\bf0.846&\bf0.918&\bf0.940&\bf0.848\\
&CsB1& 0.869&0.838 &0.035&0.999&0.999&\bf1&0.904&0.904&0.064 &0.901&0.900&0.173\\
&CsB2&\bf0.998&\bf{0.965}&\bf{0.820} &0.957& 0.979& 0.981& 0.815& 0.906& 0.845& 0.812& 0.896& 0.824\\
&rqPen &0.188 &0.194 &0.013& 0.912& 0.927& 0.997& 0.177& 0.217& 0.025& 0.095& 0.134& 0.058\\
\cline{2-5}
\cline{6-8}
\cline{9-11}
\cline{12-14}
\multirow{4}{4em}{5}&BLTQR 
 & 1& 1&{\bf 0.846} &1 &0.999 &0.999 &{\bf 1} &{\bf 0.975} &{\bf 0.916} &{\bf 1} &{\bf 0.961} &{\bf 0.919} \\
&CsB1&0.916 &1 &0.692 &0.999 &0.999 &0.999 &0.795 &0.878 &0.733 &0.804 &0.888 &0.738 \\
&CsB2&1 &1 &0.802 &0.999 &0.998 &0.999 &0.975 &0.967 &0.908 &0.972 &0.947 &0.899 \\
&rqPen &0.056 &0.037 &0.153 &0.998 &0.999 &0.996 &0.061 &0.039 &0.098 &0.060 &0.038 &0.102 \\
\bottomrule
\end{tabular*}
}
\end{table*}

Table \ref{sim-mae} presents the prediction performance for scenarios 1-5, averaged over 10 replicated datasets. The proposed model has better quantile prediction accuracy, evident from the lower check loss values compared to the alternative methods. The discrepancy of the results between the proposed model and alternative methods also becomes larger as the shape of the signals becomes more complex. This is evident from the increased gains in predictive performance under the proposed approach in other Scenarios involving slightly more complex signal shapes compared to modest gains in Scenario 1. Similar results hold for $q=0.2,0.8,$ quantiles as reported in Tables 3,6, 11, in the Supplementary Material, where the proposed approach consistently outperforms competing methods.

We conclude by reporting some specific comments about the performances of the alternative approaches. With respect to the proposed method, the alternative methods show poor out-of-sample quantile prediction, estimation, and feature selection. CsB1 and Freq. TQR report notably higher prediction errors, RE, RMSE, and lower correlation coefficients for the 2nd and 3rd visits across all scenarios, reflecting their inability to accommodate any longitudinal changes that occurred in the signals. Similarly, the vectorized approach rqPen reports poor prediction, estimation, and feature selection. The noticeably low sensitivity values indicate that rqPen cannot select significant signals, leading to high prediction error, RE, RMSE, and low correlation coefficient and MCC values. We notice that CsB2, different from all other competing methods, consistently reports high sensitivity values across all scenarios and all visits, and reaches the highest sensitivity values on the 3rd visit across all methods for scenarios 3 and 4, as evident in Table \ref{sim-fs}. This demonstrates the power of incorporating samples from all visits to estimate the visit-invariant effect $\mathcal{B}_0$. However, CsB2 reports lower specificity, F1-score, and MCC values compared to the proposed model across all scenarios, potentially due to the lack of a visit-specific term. Taking into consideration that CsB2 reports rather poor prediction and point estimation results as evident in Tables \ref{sim-fs} and \ref{sim-pe}, the high sensitivity values result from a large number of falsely selected signals, and do not indicate overall great selection performance. These findings are consistent with additional feature selection results presented in Tables 2,5, 10, in Supplementary Materials.  

To better understand the estimation and feature selection behavior of CsB2, we further evaluated the performance under an additional scenario where the differences in the signal between visits is slightly increased. The results are presented in Tables 7-8 in the  Supplementary Material. We find that with increasing noise levels, the gains in feature selection (MCC) under CsB2  for scenario 4 start to diminish across the three quantile levels, compared to the proposed method. Therefore, we believe that the proposed method shows robust advantages when at least a certain degree of difference exists for the signals at each visit.

\begin{table*}[hbt!]
\caption{Prediction results for the 5 scenarios corresponding to $q=0.5$. The best-performed methods are shown in bold.}\label{sim-mae}
\begin{tabular*}{\textwidth}{@{\extracolsep\fill}llcccc}
\toprule%
Scenarios & Methods & Visit 1 & Visit 2 & Visit 3\\
\midrule
\multirow{4}{4em}{1} & BLTQR & 0.739 &0.667 & \bf{1.317} \\ 
\cline{2-5}
& CsB TQR 1  &\bf{0.736} &\bf{0.660} &1.714  \\
\cline{2-5}
& CsB TQR 2  &2.115 & 1.492 & 3.190  \\
\cline{2-5}
& Freq. TQR &0.783 &1.594 &4.709  \\
\cline{2-5}
& rqPen  &5.271 &6.657 &6.932 \\
\midrule
\multirow{4}{4em}{2} & BLTQR & \bf{0.800} &\bf{0.684} & \bf{1.344} \\ 
\cline{2-5}
& CsB TQR 1  &0.988 & 0.962 & 4.727  \\
\cline{2-5}
& CsB TQR 2  &2.909 & 2.571 & 5.908  \\
\cline{2-5}
& Freq. TQR &3.152 &3.997 & 10.087  \\
\cline{2-5}
& rqPen  &4.672 &5.201 &9.764 \\
\midrule
\multirow{4}{4em}{3} & BLTQR & \bf{1.587} &\bf{1.543} &\bf{3.987} \\ 
\cline{2-5}
& CsB TQR 1 & 2.020 & 2.092 & 7.248  \\
\cline{2-5}
& CsB TQR 2 & 3.263 & 3.422 & 5.927  \\
\cline{2-5}
& Freq. TQR &2.814 &3.480 &10.673  \\
\cline{2-5}
& rqPen  &4.309 &4.208 &10.315 \\
\midrule
\multirow{4}{4em}{4} & BLTQR & \bf{1.906} &\bf{1.632} &\bf{2.372} \\ 
\cline{2-5}
& CsB TQR 1  &2.051 &2.060 &5.103  \\
\cline{2-5}
& CsB TQR 2  &2.844 &2.805 &4.892  \\
\cline{2-5}
& Freq. TQR &2.648 &3.174 &6.968  \\
\cline{2-5}
& rqPen & 5.299 &6.312 &8.544 \\
\midrule
\multirow{4}{4em}{5} & BLTQR
 & {\bf 1.307} & {\bf 1.193} & {\bf 1.596} \\ 
\cline{2-5}
 &CsB TQR 1  & 1.353 & 1.370 & 1.831  \\
\cline{2-5}
& CsB TQR 2  & 1.926 & 1.237 & 3.431  \\
\cline{2-5}
 & Freq. TQR &3.188 &2.833 & 4.990  \\
\cline{2-5}
& rqPen  &2.535 &3.243 &4.207 \\
\bottomrule
\end{tabular*}
\end{table*}

\subsection{Computation Times and MCMC Convergence}
The computation time for tensor rank 3 corresponding to $\mathcal{B}_0$ and $\mathcal{B}_t$ in simulations with 2D images was close to 8 mins per 100 iterations on a local computer with 1.4 GHz Processor and Memory of 8GB 2133 MHz. The corresponding computation time for $30\times 30\times 30$ 3D images is about 85 mins per 100 iterations for tensor rank 3. See Tables in Section 3 of the Supplementary Materials for additional details on the computation times. The computation times for 2-D and 3-D image analysis, along with our successful 3-D implementation, suggest that the proposed approach is scalable for practical implementation in modern imaging studies. We note that, even though the proposed approach is still scalable to larger image sizes, the computation time becomes slower with an increase in the image size as well as larger tensor ranks. If needed, computation times could be improved on computers with greater processing capability and memory, and additional speedups may be possible using more efficient computation strategies such as parallelizing the updates of the visit-specific $\mathcal{B}_t$ terms. 

A Geweke test \citep{Geweke1992PosteriorMoments} was applied to diagnose the MCMC convergence. We obtained the z-score from the Geweke test for each element of the visit-unspecific $\mathcal{B}_0$ and visit-specific $\mathcal{B}_t$. In all cases (both involving 2-D and 3-D images), we observed that over 90 percent of z-scores were within the range of (-1.96, 1.96), indicating that chains reached stationarity. MCMC trace plots of the coefficients of some of the voxels are provided in the Supplementary Material.

\subsection{Sensitivity Analysis}
We conducted a prior hyperparameter sensitivity analysis by tuning each hyperparameter while fixing other hyperparameters. Table \ref{hyperparameter-sensitivity} displays the cell-level RMSE values of the tensor coefficient matrix $\boldsymbol{B}$ of dimension $D=2$ and PARAFAC rank-\textit{R}$=3$, where $\boldsymbol{B}$ is generated from Scenario 1. The overall results reveal no strong sensitivity to the hyper-parameter choices under the selected ranges that point to a robust performance. Additionally, we evaluated the model performance as the tensor ranks were varied, and computed the corresponding DIC scores. The performance varied slightly as the tensor ranks were changed within a reasonable range, but the performance was still relatively stable.  
The results are presented in Tables 12-15 of the Supplementary Materials and suggest that the model performs reasonably well across a range of tensor ranks, producing stable and robust coefficient estimation and feature selection performance across different choices. Further, a lower choice of tensor ranks is often sufficient to yield good coefficient estimation and feature selection results, for both 2-D and 3-D cases involving sparse signals. This suggests the ability to approximate sparse signals using low rank tensors inducing model parsimony.

\subsection{Performance Under Model Mis-specification}
We consider simulations with mis-specified settings, where the data generation used normally distributed residuals instead of ALD residuals. Specifically, we generated data for 2-D images using coefficient signals corresponding to Scenarios 1-4 as described above, but using residuals generated from a standard normal distribution $N(0,1)$. The results for the $q=0.5$ quantile level, presented in Tables \ref{sim-pe-normal} and \ref{sim-fs-normal}, clearly illustrate robust point estimation and feature selection performance of the proposed BLTQR approach, even when the residuals are normally distributed and do not follow an ALD structure.  Further, the proposed approach results in improvements in coefficient estimation over the two frequentist methods that do not assume ALD errors, even under model mis-specification. This illustrates the robustness and utility of the proposed Bayesian model.

\begin{table}[hbt!]
\caption{Prior hyperparameter sensitivity analysis}\label{hyperparameter-sensitivity}
\begin{tabular}{@{}llllll@{}}
\toprule
  Hyperparameters&RMSE&Hyperparameters&RMSE&Hyperparameters&RMSE \\
\midrule
 $\alpha$ = 1/9  & 0.024  & $a_{\lambda}$ = 3 & 0.026  &$a_{\tau}$ = 1/3 & 0.024 \\
 $\alpha$= 1/6 & 0.024 &$a_{\lambda}$ = 5 & 0.028 & $a_{\tau}$ = 1/2 & 0.028\\
  $\alpha$ = 1/3& 0.022 &$a_{\lambda}$ = 7 & 0.024 &$a_{\tau}$ = 1 & 0.024  \\
  $\alpha =  3^{(-0.1)}$ & 0.027 &$a_{\lambda}$ = 9 & 0.027 &$a_{\tau}$ = 2 & 0.028 \\
\midrule
\end{tabular}
\end{table}

\begin{table*}[hbt!]
\caption{Point Estimation results of cell-level signals under model mis-specification at quantile level $q = 0.5$.}\label{sim-pe-normal}
{\footnotesize
\begin{tabular*}{\textwidth}{@{\extracolsep\fill}p{0.2cm}p{0.77cm}p{0.5cm}p{0.5cm}p{0.5cm}p{0.5cm}p{0.5cm}p{0.5cm}p{0.5cm}p{0.5cm}p{0.5cm}}
\toprule%
 \multirow{2}{4em}{Scenarios} &  &  \multicolumn{3}{c}{RE} & \multicolumn{3}{c}{RMSE}& \multicolumn{3}{c}{Corr. Coef.}\\
\cline{3-5}
\cline{6-8}
\cline{9-11}
& & v1&v2&v3&v1&v2&v3&v1&v2&v3\\
\midrule
\multirow{5}{4em}{1} &BLTQR &0.137 & 0.108&0.245 & 0.021 & 0.018& 0.059& 0.997& 0.998& 0.994\\
&Freq. TQR &0.330 &0.170 &0.927 &0.043 &0.026 &0.205 &0.986 &0.995 &0.837 \\
&rqPen & 0.995&0.995 & 1.025&0.297 &0.297 & 0.389&0.126 &0.127 &0.047 \\
\midrule
\multirow{5}{4em}{2} &BLTQR &0.243 &0.185 &0.242 &0.026  &0.021 &0.058 &0.993 &0.997 &0.993\\
&Freq. TQR &1.670 & 1.700 &2.608 &0.157 &0.170 &0.537 & 0.760 &0.734 & 0.310\\
&rqPen &1.107 &1.344 &1.001 &0.221 &0.271 &0.442 &0.371 &0.177 &0.171 \\
\midrule
\multirow{5}{4em}{3}&BLTQR &0.646 &0.648 &0.936 &0.068 &0.074 &0.203 &0.951 &0.943 &0.890\\
&Freq. TQR &1.414 &1.339 &2.799 & 0.137& 0.147& 0.503&0.802 &0.773 &0.340\\
&rqPen &1 &1.161 &1.005 &0.225 &0.224 &0.449 &0.261 &0.302 &.002 \\
\midrule
\multirow{5}{4em}{4}&BLTQR &0.454 &0.333 &0.397 & 0.085 & 0.082 &0.102 &0.955 &0.968 & 0.960\\
&Freq. TQR &0.750 &0.729 &1.544 &0.123 &0.141 &0.345 &0.898 &0.894 &0.562 \\
&rqPen &1.288 &1.019 & 1.024&0.312 &0.334 & 0.395&0.176 &0.086 &0.209 \\
\bottomrule
\end{tabular*}
}
\end{table*}

\begin{table*}[hbt!]
{\footnotesize
\caption{Feature selection under model mis-specification at quantile level $q = 0.5$}\label{sim-fs-normal}
\begin{tabular*}{\textwidth}{@{\extracolsep\fill}p{0.2cm}p{0.77cm}p{0.5cm}p{0.5cm}p{0.5cm}p{0.5cm}p{0.5cm}p{0.5cm}p{0.5cm}p{0.5cm}p{0.5cm}p{0.5cm}p{0.5cm}p{0.5cm}}
\toprule%
 \multirow{2}{4em}{Scenarios} & &  \multicolumn{3}{c}{Sens.} & \multicolumn{3}{c}{Spec.}& \multicolumn{3}{c}{F1-score}& \multicolumn{3}{c}{MCC}\\
\cline{3-5}
\cline{6-8}
\cline{9-11}
\cline{12-14}
& &v1&v2&v3&v1&v2&v3&v1&v2&v3&v1&v2&v3\\
\midrule
 \multirow{2}{4em}{1}&BLTQR &1& 1& 1& 1& 1& 1& 1 & 1& 1& 1 & 1& 1\\
&rqPen &0.038 &0.038 &0.021& 0.995& 0.996& 0.992& 0.071& 0.071& 0.038& 0.114& 0.114& 0.040\\
\cline{2-5}
\cline{6-8}
\cline{9-11}
\cline{12-14}
 \multirow{2}{4em}{2}&BLTQR & 1& 1& 1& 0.999& 1& 1&0.999 & 1& 1& 0.999& 1& 1\\
&rqPen &0.384 &0.296 &0.095& 0.942& 0.912& 0.991& 0.322& 0.228& 0.162& 0.281& 0.169& 0.196\\
\cline{2-5}
\cline{6-8}
\cline{9-11}
\cline{12-14}
 \multirow{2}{4em}{3}&BLTQR &0.997& 0.962& 0.748 &0.995& 0.996& 0.999& 0.954& 0.950& 0.775& 0.953& 0.948& 0.793\\
&rqPen &0.218 &0.338 &0.006& 0.951& 0.939& 0.999& 0.195& 0.278& 0.012& 0.148& 0.235& 0.048\\
\cline{2-5}
\cline{6-8}
\cline{9-11}
\cline{12-14}
\multirow{2}{4em}{4}&BLTQR &0.987& 0.966& 0.948 &0.996& 0.999& 0.999& 0.974& 0.979& 0.971& 0.971& 0.977& 0.967 \\
&rqPen &0.248 &0.035 &0.123& 0.900& 0.990& 0.959& 0.214& 0.063& 0.176& 0.131& 0.070& 0.124\\
\bottomrule
\end{tabular*}
}
\end{table*}

\section{ADNI-1 Data Analysis}\label{sec-adni}
We analyze longitudinal data from the Alzheimer’s Disease Neuroimaging Initiative (ADNI) 1 study collected at baseline, 6-month visit, and 12-month visit.  ADNI is a large-scale, longitudinal study that was launched in 2004 to explore the progression of Alzheimer's disease (AD) through neuroimaging and biomarker analysis \citep{weiner2013alzheimer}.  More detailed information for the ADNI-1 sample included for our analysis is provided in Table 16 (Section 4) in the Supplementary Material. The revised analyses using 3D ROIs include MCI subjects with sample sizes of 403, 311, and 321 across three visits, as well as a separate analysis for AD subjects with sample sizes of 188, 133, and 135 across three visits. Out of these, 308 MCI and 133 AD subjects have no missing visits, while other subjects have one or more missing visits.

\subsection{Data Source and Pre-processing}
The T1-weighted MRI images were processed using the Advanced Normalization Tools (ANTs) registration pipeline \citep{tustison2014large}. This involved registering all images to a template to ensure consistent normalization of brain regions across participants. The population-based template was created using data from 52 normal control participants in ADNI 1, originally developed by the ANTs group \citep{tustison2019longitudinal}. Notably, the ANTs pipeline includes the N4 bias correction step to address intensity variations and standardize intensity across samples \citep{tustison2010n4itk}. It also uses a symmetric diffeomorphic image registration algorithm for spatial normalization, aligning each T1 image with a brain template to enable spatial comparability \citep{avants2008symmetric}. Afterward, the pipeline used the processed brain images, estimated brain masks, and template tissue labels to perform 6-tissue Atropos segmentation, creating tissue masks for cerebrospinal fluid (CSF), gray matter (GM), white matter (WM), deep gray matter (DGM), brain stem, and cerebellum. Finally, cortical thickness was measured using the DiReCT algorithm. The 3D cortical thickness images of dimension $256 \times 256 \times 170$ were cropped to dimension $138 \times 163 \times 117$ to remove the zero voxels around the images, and then downsampled to a resolution of $48 \times 48 \times 48$ using R package \texttt{ANTsRCore}. The downsampled cortical thickness images were harmonized using the ComBat approach \citep{fortin2018harmonization} to remove scanner-specific batch effects that can inflate variability and cause confounding, if not accounted for. The NeuroComBat package in R was used for harmonization.

Finally, the brain was parcellated into 83 regions of interest (ROIs) based on the DKT atlas \citep{Desikan2012CorticalLabeling, Lawrence2021StandardizingParcellations}, and the analysis was performed separately for each 3-D ROI. We used a cortical mask to handle the sparsity in the cortical thickness images when implementing the model, where the mask was pre-specified using the distribution of cortical regions in the brain image data. Posterior distributions were computed by ignoring the voxels lying outside the cortical mask, and posterior estimates of the tensor coefficients for any voxel lying outside the cortical mask were thresholded to be zero.

\begin{figure}[h]
\centering
\includegraphics[width=1\textwidth]{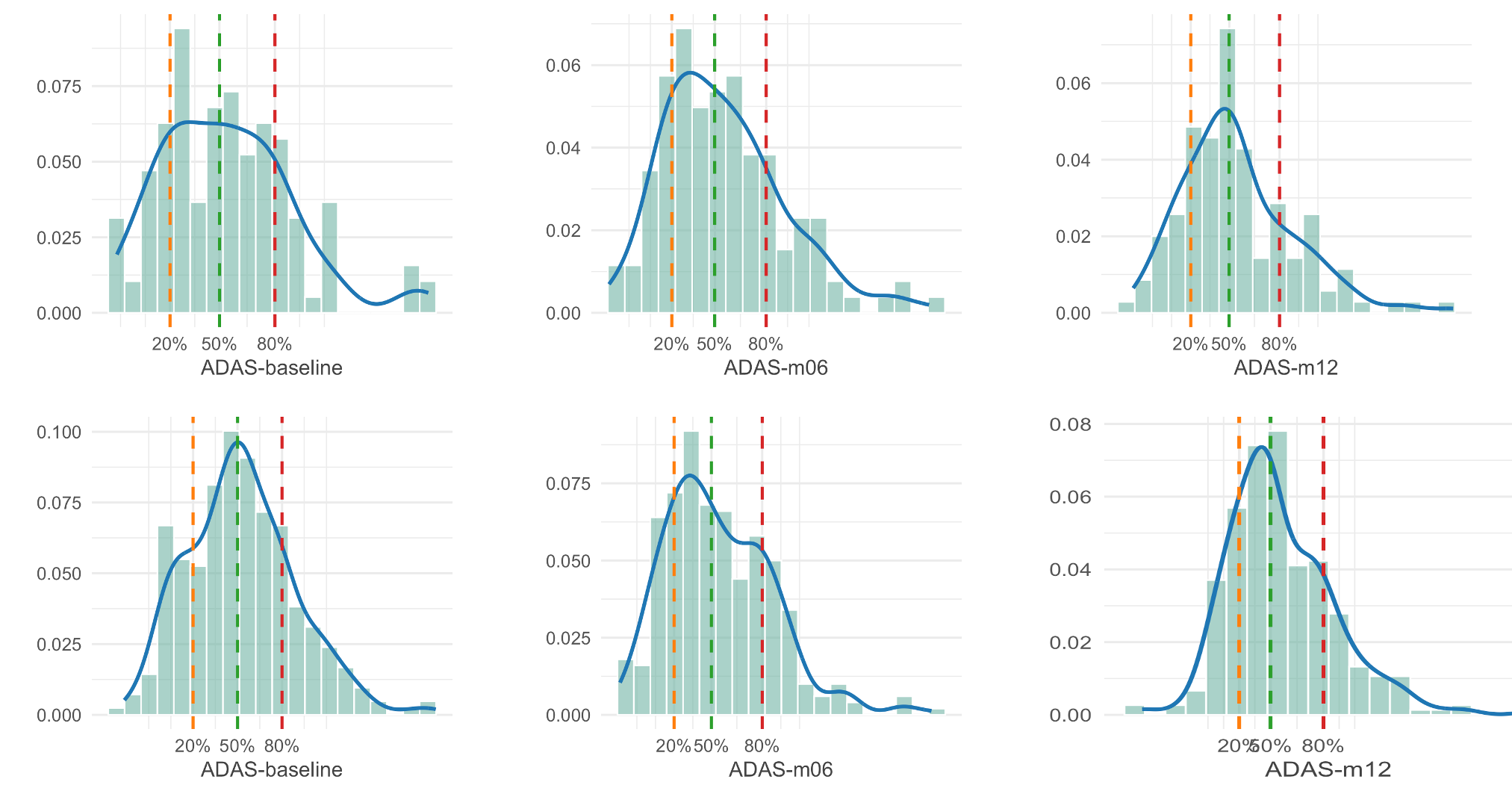}
\caption{From left to right: Top panel: The distribution of observed ADAS scores measured at baseline, 6-month visit, and 12-month visit for AD cohort; Bottom panel: The distribution of observed ADAS scores measured at baseline, 6-month visit, and 12-month visit for MCI cohort. The test scores have skewed distributions. The quantiles are labeled on the x-axis.}
\label{adas-distribution}
\end{figure}

\subsection{Analysis Outline}
Our goal is to predict cognitive impairment based on cortical thickness data (derived from T1-MRI scans) along with additional covariates such as baseline age, gender, years of education, APOE4 allele (0,1,2), and intracranial volume (ICV), and infer brain regions that are significantly associated with cognitive decline. To achieve this, we applied the proposed approach to model the  Alzheimer's Disease Assessment Scale (ADAS) scores, using longitudinal data from the ADNI-1 study. The ADAS score assesses decline in memory, language, and praxis in Alzheimer's disease (AD), and is frequently used in research and clinical settings to monitor Alzheimer's progression and to evaluate the impact of treatments. Figure \ref{adas-distribution} visualizes the ADAS scores over three visits, 
highlighting a noticeable rightward shift in the quantiles for later visits.  This is indeed corroborated by the demographics Table 16 in Supplementary Materials (Section 4), which shows an increase in the 80th quantiles of the ADAS scores over time for the MCI cohort, suggesting a decline in cognitive ability over the course of the visits.
Previous studies addressed the importance of including both cross-sectional and longitudinal covariate effects along with longitudinal brain data, especially for the effect of age \citep{sorensen2021recipe,neuhaus1998between}. We account for this by including an interaction term between age and time from baseline in the regression model (i.e. include $Age*\mathcal{T}_{it}$ on the right hand side of the equation (\ref{model})), in addition to the other covariates already included for analysis. We analyzed the AD and MCI cohorts separately, and evaluated the prediction performance for each group. Specifically, we performed the prediction task by including all data from the 1st and 2nd visits and randomly selecting 70\% of the 3rd visit as the training set, using the remaining 30\% of the data from the 3rd visit as the testing set to evaluate prediction performance. We compared results from the proposed method with (i) CsB1 and CsB2 and (ii) Freq. TQR, as previously described in Section \ref{sim-results}. Prediction results were averaged over 10 replicates and reported separately for each ROI.  For feature selection, we report significant voxels associated with quantiles of the cognitive score based on 90\% point-wise credible intervals. Unfortunately, most significant effects did not survive multiplicity adjustments due to smaller effect sizes and noise in the brain images for $\alpha=0.1$. However, we did get a small number of significant voxels after multiplicity adjustment for smaller choices of $\alpha$(results not reported).

\subsection{Results}

\begin{figure}[hbt!]%
\centering
\includegraphics[width=1\textwidth]{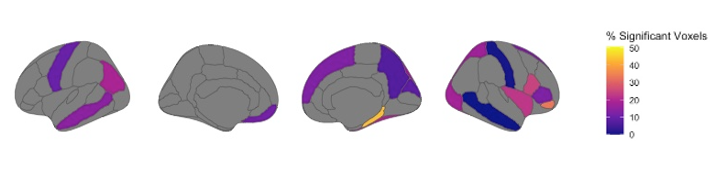}
\caption{Proportion of significant voxels in selected ROIs in the DKT atlas at the 0.5 quantile, for some select ROIs  having the highest number of significant voxels. The remaining ROIs are depicted in gray and correspond to regions with limited proportion of significant voxels.}\label{ADNI_mci}
\end{figure}

We fit the model corresponding to each ROI (under the DKT atlas) based on tensor ranks ranging from 2-4 for $\mathcal{B}_0$ and $\mathcal{B}_t$, and chose the optimal rank based on their DIC values. The optimal tensor rank may vary across different ROIs. In several cases,  the optimal tensor rank was 3 for $\mathcal{B}_0$ and 2 for $\mathcal{B}_t$ under the DIC score. The DIC values corresponding to varying tensor rank choices under some selected ROIs for the ADNI data analysis are reported in Table 23 in the Supplementary Materials. We ran 8000 iterations, with 4000 burn-ins. According to the Geweke test \citep{Geweke1992PosteriorMoments}, around 75\% of the chains reached stationary.

{ables \ref{adni-mci-mae} and \ref{adni-ad-mae} show the prediction performance for AD and MCI subjects, respectively, at quantile levels 0.2, 0.5, and 0.8 for a subset of ROIs that have the highest number of significant voxels after feature selection, with standard errors reported in parentheses. Prediction results for all ROIs are presented in Tables 17-22 in Section 4 of the Supplementary Materials. The proposed method shows lower check loss values than all competing methods across all ROIs and for the three quantile levels considered. The improvement in results under the proposed method holds for both the MCI and AD cohorts. In particular, the proposed method performs consistently better than CsB1 and CsB2, demonstrating the advantages of simultaneously including the visit-specific effect $\mathcal{B}_t$, along with shared information across visits via the visit-invariant term $\mathcal{B}_0$. Moreover, the Freq. TQR has the poorest performance overall, consistently across all scenarios. Further all approaches have a less accurate prediction in the AD cohort compared to the MCI cohort due to the smaller sample size in the AD cohort. However, the relative deterioration in prediction for the AD cohort under the Freq. TQR approach is much worse compared to the Bayesian methods. We also observe some fluctuations in performance across different ROIs, which may be attributed to variations in the proportion of cortical tissue contained within each ROI, as well as the differing contributions of specific brain regions to cognitive function. In summary, the prediction results under the proposed approach consistently outperform the competing methods, showcasing the strength of the tensor-based model incorporating visit-specific and visit-invariant terms for quantile regression based on neuroimaging data. 

\subsubsection{Brain region discovery} We  limit our findings to the MCI cohort, since the analysis of the AD cohort produced a small number of significant brain regions related to cognitive performance. The proposed approach discovers several ROIs (under the DKT atlas) with a large number of significant voxels inferred for the MCI analysis corresponding to $q=0.5$. This set of ROIs is reported in Table \ref{tab:MCIvoxels}, and also visualized in Figure \ref{ADNI_mci}. The reported ROIs cover areas belonging to the precuneus and inferior/superior parietal cortex (ROIs 20, 60, 18), temporal association/ventral temporal regions (30, 25), occipital cortex (ROI 23) and sensorimotor regions (ROI 57). 

ROI 18 is located in the right superior parietal association cortex near the precuneus–SPL (superior parietal lobule) neighborhood. Posterior parietal cortical thinning is frequently observed in MCI and represents a robust structural marker of prodromal Alzheimer-related neurodegeneration, particularly when co-occurring with other cortical signature regions \citep{Singh2006CorticalThinning, Dickerson2009, du2007different}. These posterior association cortices, including superior and inferior parietal territories, exhibit early vulnerability to AD pathophysiology \citep{Buckner2005}. ROI 20 is located in the right precuneus (default mode network) and close to the posterior medial cortex, which is a major default mode network hub. Cortical atrophy in the precuneus is detectable at the very earliest stages of the disease \citep{Buckner2005} is a robust feature of the AD cortical signature as it relates to clinical severity \citep{Dickerson2009}. Moreover, significant reductions in glucose metabolism are noted in the precuneus that often correlate with the severity of cognitive impairment as measured by MMSE scores \citep{Buckner2005}. ROI 60 is located in left inferior parietal association cortex (within the default mode network)  that includes the angular gyrus and supramarginal gyrus, which are a core component of the cortical signature of AD and is significantly impacted across the stages of MCI and AD \cite{Dickerson2009}. Quantitative analyses identify significant structural decay in these regions at the earliest clinical stages; for example, the angular gyrus exhibits approximately 4.46\% thinning in incipient AD, which accelerates to 9.60\% in mild AD \citep{Dickerson2009}. Singh et al. \citep{Singh2006CorticalThinning} reported that these atrophic patterns are systematically more pronounced in the left hemisphere and they are informative of the transition from healthy aging to clinical dementia. This thinning is directly linked to cognitive impairment severity \citep{du2007different}. Other ROIs with a large number of significant voxels include ROI 25 that lies in the right fusiform gyrus within the ventral temporal fusiform cortex, and ROI 30 that lies in the right lateral temporal cortex. Existing literature illustrates that ventral temporal regions, including the inferior temporal and lateral occipitotemporal (fusiform) gyri, exhibit substantial atrophy in MCI and mild AD \citep{Dickerson2009}. 

Our analysis also indicated some important regions with a large number of significant voxels that may indicate novel findings and potentially require additional validation. Our analysis discovered large number of significant voxels in ROI 23 that is located in the right lateral occipital association cortex near the occipital pole. In typical MCI and typical amnestic AD, occipital cortex is often less affected early compared with posterior medial and temporo-parietal regions \citep{Braak1991, Dickerson2009}. However, occipital involvement is more plausible in atypical/posterior cortical atrophy (PCA) presentations in AD, where occipito-parietal degeneration is more prominent \citep{Crutch2012}. We also identified a large number of significant voxels in ROI 57 that is situated in the left precentral gyrus (primary motor cortex). Although primary cortices including motor regions are relatively preserved in early disease stages compared with temporo-parietal and posterior medial association areas \citep{Braak1991, Dickerson2009, Salat2004}, emerging evidence suggests that motor cortex changes may arise through network-level degeneration, reflecting the progressive disintegration of large-scale functional systems rather than isolated regional vulnerability \citep{Seeley2009NeurodegenerativeNetworks, Salat2004, Brier2012LossFunctionalConnectivity}.
For example, \citep{Seeley2009NeurodegenerativeNetworks}
demonstrated that motor cortex changes appear as part of the disintegration of a dorsal sensorimotor association network. Further, \citep{Brier2012LossFunctionalConnectivity} found that as the disease progresses to mild dementia, the Sensory-Motor Network (SMN) exhibits a significant loss of intranetwork functional correlations. They observed that anticorrelations between the default mode network (DMN) and motor regions are reduced even at the mild symptomatic stage, suggesting a mechanism for pathology to spread from the DMN to the rest of the brain.

\begin{table}[htbp]
\centering
\caption{Prediction performance for MCI subjects from 3D ADNI images of a subset of ROIs with the largest number of significant voxels, at quantile level 0.2, 0.5 and 0.8.}\label{adni-mci-mae}
{\scriptsize
\begin{tabular}{c cccc cccc cccc}
\toprule
 & \multicolumn{4}{c}{$q = 0.2$} 
 & \multicolumn{4}{c}{$q = 0.5$} 
 & \multicolumn{4}{c}{$q = 0.8$} \\
\cmidrule(lr){2-5}\cmidrule(lr){6-9}\cmidrule(lr){10-13}
ROI 
& Proposed & CsB1 & CsB2 & Freq.\ TQR
& Proposed & CsB1 & CsB2 & Freq.\ TQR
& Proposed & CsB1 & CsB2 & Freq.\ TQR \\
\midrule
57 & \textbf{1.294} & 1.605 & 1.478 & 2.931 & \textbf{2.020} & 2.355 & 2.150 & 3.625 & \textbf{1.654} & 1.866 & 1.740 & 2.953 \\
6  & \textbf{1.337} & 1.467 & 1.483 & 3.013 & \textbf{2.164} & 2.447 & 2.398 & 3.886 & \textbf{1.744} & 2.035 & 1.856 & 3.014 \\
23 & \textbf{1.328} & 1.554 & 1.484 & 2.757 & \textbf{2.120} & 2.381 & 2.277 & 3.738 & \textbf{1.742} & 2.105 & 1.869 & 2.949 \\
8  & \textbf{1.267} & 1.402 & 1.391 & 3.421 & \textbf{2.072} & 2.291 & 2.280 & 3.751 & \textbf{1.622} & 2.034 & 1.709 & 3.177 \\
18 & \textbf{1.330} & 1.385 & 1.349 & 2.938 & \textbf{2.075} & 2.445 & 2.378 & 3.851 & \textbf{1.617} & 1.942 & 1.952 & 3.088 \\
7  & \textbf{1.316} & 1.488 & 1.463 & 3.157 & \textbf{2.109} & 2.380 & 2.282 & 3.647 & \textbf{1.703} & 2.271 & 1.937 & 3.123 \\
20 & \textbf{1.301} & 1.501 & 1.432 & 3.075 & \textbf{2.114} & 2.354 & 2.278 & 3.694 & \textbf{1.731} & 2.075 & 1.739 & 3.005 \\
25 & \textbf{1.268} & 1.614 & 1.440 & 2.710 & \textbf{2.080} & 2.265 & 2.264 & 3.626 & \textbf{1.655} & 1.834 & 1.740 & 2.887 \\
30 & \textbf{1.295} & 1.542 & 1.415 & 2.987 & \textbf{1.984} & 2.542 & 2.397 & 3.838 & \textbf{1.659} & 1.930 & 1.691 & 2.761 \\
71 & \textbf{1.303} & 1.764 & 1.456 & 2.903 & \textbf{2.115} & 2.416 & 2.351 & 3.491 & \textbf{1.720} & 1.932 & 1.837 & 2.661 \\
21 & \textbf{1.307} & 1.527 & 1.415 & 2.934 & \textbf{2.172} & 2.490 & 2.389 & 3.871 & \textbf{1.793} & 1.943 & 1.890 & 3.157 \\
60 & \textbf{1.318} & 1.508 & 1.399 & 2.599 & \textbf{2.032} & 2.360 & 2.225 & 3.885 & \textbf{1.615} & 1.774 & 1.643 & 2.869 \\
62 & \textbf{1.308} & 1.730 & 1.543 & 2.958 & \textbf{2.202} & 2.579 & 2.457 & 3.468 & \textbf{1.899} & 2.088 & 1.959 & 3.379 \\
40 & \textbf{1.306} & 1.591 & 1.472 & 2.966 & \textbf{2.126} & 2.441 & 2.316 & 3.649 & \textbf{1.731} & 1.982 & 1.765 & 2.987 \\
51 & \textbf{1.284} & 1.743 & 1.430 & 2.739 & \textbf{2.378} & 2.476 & 2.349 & 3.842 & \textbf{1.753} & 1.910 & 1.853 & 2.983 \\
5  & \textbf{1.399} & 1.475 & 1.468 & 2.978 & \textbf{2.222} & 2.523 & 2.543 & 3.920 & \textbf{1.799} & 1.942 & 1.899 & 2.998 \\
55 & \textbf{1.331} & 1.686 & 1.506 & 2.859 & \textbf{2.101} & 2.509 & 2.344 & 3.621 & \textbf{1.678} & 1.917 & 1.860 & 3.035 \\
\bottomrule
\end{tabular}
}
\end{table}

\begin{table}[htbp]
\centering
\caption{Prediction performance for AD subjects from 3D ADNI images of a subset of ROIs with the largest number of significant voxels, at quantile level 0.2, 0.5 and 0.8.}\label{adni-ad-mae}
\label{tab:ad_reduced_allq_final}
{\scriptsize
\begin{tabular}{c cccc cccc cccc}
\toprule
& \multicolumn{4}{c}{$q = 0.2$}
& \multicolumn{4}{c}{$q = 0.5$}
& \multicolumn{4}{c}{$q = 0.8$} \\
\cmidrule(lr){2-5}\cmidrule(lr){6-9}\cmidrule(lr){10-13}
ROI
& Proposed & CsB1 & CsB2 & Freq.\ TQR
& Proposed & CsB1 & CsB2 & Freq.\ TQR
& Proposed & CsB1 & CsB2 & Freq.\ TQR \\
\midrule
62 & \textbf{2.332} & 2.424 & 2.369 & 5.780
   & \textbf{3.470} & 3.753 & 3.645 & 5.611
   & \textbf{2.770} & 2.910 & 2.789 & 5.418 \\

58 & \textbf{2.184} & 2.337 & 2.314 & 5.829
   & \textbf{3.413} & 3.941 & 3.880 & 6.032
   & \textbf{2.748} & 3.063 & 2.960 & 5.576 \\

82 & \textbf{2.137} & 2.248 & 2.274 & 5.799
   & \textbf{3.450} & 3.991 & 3.615 & 5.970
   & \textbf{2.528} & 2.727 & 2.604 & 5.494 \\

43 & \textbf{2.240} & 2.373 & 2.287 & 5.874
   & \textbf{3.688} & 3.758 & 3.752 & 5.785
   & \textbf{2.329} & 3.016 & 2.867 & 5.834 \\

80 & \textbf{2.222} & 2.396 & 2.322 & 5.802
   & \textbf{3.330} & 3.927 & 3.782 & 5.908
   & \textbf{2.764} & 3.035 & 2.889 & 5.556 \\

74 & \textbf{2.136} & 2.228 & 2.225 & 5.429
   & \textbf{3.519} & 3.764 & 3.685 & 5.687
   & \textbf{2.462} & 2.893 & 2.579 & 5.188 \\

46 & \textbf{2.165} & 2.319 & 2.191 & 5.982
   & \textbf{3.158} & 3.734 & 3.446 & 5.505
   & \textbf{2.599} & 2.987 & 2.701 & 5.805 \\

28 & \textbf{2.145} & 2.388 & 2.415 & 5.743
   & \textbf{3.383} & 3.897 & 3.815 & 5.963
   & \textbf{2.560} & 2.920 & 2.766 & 5.156 \\

5  & \textbf{2.235} & 2.370 & 2.355 & 5.808
   & \textbf{3.547} & 3.562 & 3.553 & 5.850
   & \textbf{2.789} & 2.932 & 2.851 & 5.912 \\

15 & \textbf{2.284} & 2.479 & 2.431 & 5.491
   & \textbf{3.692} & 3.803 & 3.854 & 6.061
   & \textbf{2.791} & 2.971 & 2.866 & 5.262 \\

\bottomrule
\end{tabular}
}
\end{table}

\begin{table*}[hbt]
\caption{ROIs with a large number of significant voxels under the MCI analysis at q = 0.5.}\label{tab:MCIvoxels}
\begin{tabular*}{\textwidth}{@{\extracolsep\fill}lllc}
\toprule%
 ROI Index & Brain Region &\# voxel selected & \% of ROI  \\
\midrule
57 & L. Precentral Gyrus & 819& 0.142 \\
\midrule
6 & R. Insular gyrus &210&0.119\\
\midrule
23 & R. lateral occipital cortex &353&0.114\\
\midrule
8 & R. superior frontal gyrus & 478&0.113\\
\midrule
18 & R. dorsal parietal cortex &425&0.094\\
\midrule
7 & R. frontal pole &351&0.093\\
\midrule
20 & R. precuneus &192&0.083\\
\midrule
25 & R. fusiform gyrus &187&0.07\\
\midrule
30 & R. right middle temporal gyrus &201&0.07\\
\midrule
71 & L. middle temporal gyrus &176&0.07\\
\midrule
21 &R. cuneus gyrus &41&0.062\\
\midrule
60 & L. inferior parietal cortex &295&0.06\\
\midrule
 62 & L. medial occipital cortex &62&0.057\\
\midrule
40 & R. parahippocampal gyrus &48&0.056\\
\midrule
51 &  L. peri-central sensorimotor cortex &472&0.056\\
\midrule
 5 & R. inferior frontal gyrus &42&0.055\\
\bottomrule
\end{tabular*}
\end{table*}

\section{Conclusion}

In this study, we have proposed a novel Bayesian tensor quantile regression for longitudinal high-dimensional imaging data. The proposed approach makes an important contribution to the Bayesian quantile regression literature by including high-dimensional spatial functions as covariates in longitudinal settings, for which there is limited literature.  A quantile-based modeling approach allows to investigate how the imaging covariates are associated with different parts of the distribution of cognitive performance scores, which provides potential neurobiological biomarkers for downstream analysis, as demonstrated via the analysis of ADNI-1 data. Moreover, the ability to incorporate information across multiple visits allows us 
to study the overall longitudinal trajectory of disease progression. 
The proposed approach is flexible in terms of allowing one to borrow information across visits via a visit-invariant term, and to simultaneously include sparse tensor terms the control visit-specific effects contributing to the overall longitudinal trajectory. The proposed approach naturally accounts for missing longitudinal visits, allowing for varying number of visits across samples. 

We employed PARAFAC decompositions on the tensor coefficients, to preserve the spatial configuration and avoid the curse of dimensionality. We incorporated multiway shrinkage priors to model the visit-invariant tensor coefficients and variable selection priors on the tensor margins of the visit-specific effects. For posterior inference, we developed a computationally efficient Markov chain Monte Carlo sampling algorithm. Our synthetic experiments involving 2-D and 3-D images showcase the accuracy and scalability of the approach. Finally, the 3-D analysis of ADNI-1 dataset reveals a superior ability of the proposed approach to predict cognitive scores, and infer brain regions with significant effects on different parts of the cognitive  score distribution. We used our analysis to identify  brain regions that significantly contribute to worse cognitive outcomes that has considerable translational importance.

 Our work is reminiscent of recently proposed 4-D tensor approaches for modeling spatiotemporal objects that involve spatiotemporal varying coefficients (STVC) constructions \citep{Lei2025, Niyogi2024}. These approaches model the regression coefficients corresponding to spatiotemporal images as an outer product of tensor margins, where one tensor margin represents temporal variations, and the remaining tensor margins capture the spatial aspects. Therefore, these frameworks encode the temporal effects as a multiplicative term, which are potentially more suited for scenarios with smoothly varying global longitudinal changes across spatial locations and induce model parsimony. However, such a construction may not be flexible enough to model 
localized voxel-wise longitudinal differences, which is our focus. On the contrary, the proposed approach is designed to improve the ability to model localized voxel-wise differences across visits by separating the signals into an additive tensor modeling framework comprising a visit-invariant component ($\mathcal{B}_0$) and visit-specific components ($\mathcal{B}_t$). The spike and slab shrinkage priors imposed on the  tensor margin elements is designed to model localized differences (or lack thereof) across visits. Simultaneously, the proposed model can learn global signals via the time-invariant term ($\mathcal{B}_0$) that is estimated by combining information across visits. The ability to learn localized signals via additive tensors comes at the cost of reduced model parsimony compared to 4-dimensional tensors; however our numerical examples illustrate the applicability of the proposed approach to high dimensional images.

Another consideration is the assumption of similar visit timing across samples that is motivated by the  design of ADNI study that collected data at regular intervals for all samples \citep{Jack2008ADNIMRIMethods}, but may not be applicable to other studies with irregular visit timings. To accommodate such scenarios, the proposed model may require additional generalizations that will be the focus of future work. Additionally, this study analyzes AD and MCI cohorts separately, due to potential heterogeneity across these cohorts. Future research will address this limitation by combining AD and MCI cohorts, and explicitly accounting for heterogeneity across samples due to disease status and other factors  relevant for AD research.

\bibliography{bibliography}       

\end{document}